\documentclass[aps,pra,twocolumn,color,longbibliography]{revtex4-2}
\usepackage{amsmath,amsfonts,amssymb}
\usepackage{bm}
\usepackage{graphicx}
\usepackage{array,color}
\usepackage{braket}
\usepackage{xcolor}

\begin{document}
	
\title{Phase separation and metastability in a mixture of spin-1 and spin-2\\ Bose-Einstein condensates}

\author{Uyen Ngoc Le}
\affiliation{Department of Engineering Science, University of Electro-Communications, Tokyo 182-8585, Japan}

\author{Hieu Binh Le}
\affiliation{Department of Engineering Science, University of Electro-Communications, Tokyo 182-8585, Japan}

\author{Hiroki Saito}
\affiliation{Department of Engineering Science, University of Electro-Communications, Tokyo 182-8585, Japan}

\date{\today}
\begin{abstract}
We investigate the ground state and dynamics of a mixture of spin-1
and spin-2 Bose-Einstein condensates of ${}^{87}{\rm{Rb}}$ atoms. For
the experimentally measured interaction coefficients, the ground state
exhibits phase separation between the spin-1 ferromagnetic state and
the spin-2 nematic state. At the interface between them, a partially polarized spin state emerges. The uniformly mixed state of the spin-1 polar state and spin-2 biaxial nematic state is metastable, and the phase separation via nucleation can be triggered by a local perturbation.
\end{abstract}

\maketitle

\section{Introduction}
\label{sec:first}
Macroscopic coherent matter waves with internal degrees of freedom,
such as superfluid ${}^{3}{\rm{He}}$ \cite{vollhardt2013superfluid},
\textit{p}-wave and \textit{d}-wave superconductors
\cite{norman2011challenge}, and spinor Bose-Einstein condensates
(BECs) of atomic gases \cite{Ohmi1998, Ho1998, Law1998,
  Ho1998,Stamper1998, stenger1998spin}, have attracted great interest
because of their rich variety of quantum features. In particular, the
experimental systems of ultracold atoms are highly controllable, and
various studies on the spinor BECs, including studies on their
spin-mixing dynamics \cite{stenger1998spin, Chang2004, Chang2005,
  Schmaljohann2004}, topological excitation \cite{al2001, Leslie2009,
  Choi2012, ray2014, Seo2015, hall2016tying}, and spin textures \cite
{sadler2006, Eto2014}, have been reported.

The ground-state phases of the spinor BECs depend on the spin-dependent interactions between atoms.
For the spin-1 BEC, there are two types of ground states: the polar
and ferromagnetic states \cite{Ohmi1998, Ho1998, Law1998, Ho1998}. The
phase diagram is more complicated for the spin-2 BEC, which includes the cyclic phase \cite{Ciobanu2000, Koashi2000, Ueda2002, gorlitz2003, Kuwamoto2004, Barnett2006}. The spin-3 BEC exhibits eleven ground-state phases \cite{ho2006, santos2006, makela2007}. An external magnetic field modifies the phase diagrams \cite{stenger1998spin}.

Mixtures of two or more spinor BECs can enrich the physics
further. The phase diagrams and many-body properties of a binary
mixture of spin-1 BECs have been theoretically studied
\cite{luo2007bose, xu2009binary, shi2010ground, xu2010quantum,
  xu2010spontaneously, zhang2010atomic, shi2011three,
  zhang2011interspecies, xu2011quantum, xu2012quantum,
  zhang2015fragmentation, chen2018resonant, he2019, he2019_2,
  jie2021laser} and experimentally realized for spin-1
${}^{87}{\rm{Rb}}$ and ${}^{23}{\rm{Na}}$ atoms
\cite{li2015coherent}. Recently, the research was extended to a
three-component mixture of spin-1 BECs
\cite{liu2017,he2020,he2022}. The ground-state phase diagrams of a
mixture of spin-1 and spin-2 BECs, including their broken-axisymmetry
phases \cite{irikura2018}, have also been reported. The dynamics of a
spin-1 BEC interacting with a spin-2 BEC have been observed for ${}^{87}{\rm{Rb}}$ atoms \cite{eto2018}.

Most of the previous studies on mixtures of spinor BECs have been
restricted to the single-mode approximation (SMA), where the spatial
degrees of freedom are frozen \cite{xu2009binary, shi2010ground,
  chen2018resonant, zhang2010atomic, shi2011three,
  zhang2011interspecies, xu2011quantum, xu2012quantum,
  zhang2015fragmentation, xu2010quantum, jie2021laser}. For a mixture
of spin-1 and spin-2 BECs, Ref.~\cite{irikura2018} also relied on the
SMA, which showed that the ground state of a 1:1 mixture for
${}^{87}{\rm{Rb}}$ atoms is the polar state for spin-1 and the biaxial
nematic state for spin-2. However, the possibility arises that phase separation occurs in a system much larger than the spin healing length, which cannot be captured by the SMA. The purpose of the present paper is to explore the possibility of phase separation in the spin-1 and spin-2 BECs of ${}^{87}{\rm{Rb}}$ atoms.

In this paper, using mean-field theory, we show two main
results. First, the ground state of the mixture of spin-1 and spin-2
BECs of ${}^{87}{\rm{Rb}}$ atoms exhibits phase separation into the
two phases: the ferromagnetic state for spin-1 and the nematic state for spin-2. In the interface layer between these two phases, a distinct phase emerges in which both components have partial magnetizations into the opposite directions. Second, the uniformly mixed state of the spin-1 polar state and spin-2 biaxial nematic state can be metastable. If a local perturbation is imparted to this mixture, the phase separation is triggered, which extends over the whole space. We will show that the phase separation via nucleation can be observed even in the presence of the inelastic collisional decay of spin-2 ${}^{87}{\rm{Rb}}$ atoms.

This paper is organized as follows. Section \ref{sec:second} provides
a formulation of the problem. Section \ref{sec:third} reveals that the
ground state exhibits phase separation. Section~\ref{sec:fourth} shows
that there exists a uniformly mixed metastable state and demonstrates the dynamics of phase separation via nucleation. Section \ref{sec:fifth} proposes an experiment to observe the phase separation via nucleation, and Sec.~\ref{sec:concl} summarizes the results.

\section{Formulation of the problem}
\label{sec:second}
In the mean-field approximation at zero temperature, the spin-1 and spin-2 BECs can be described by the macroscopic wave function $\psi _m^{(f)}(\bm{r})$, where $f = 1,2$ is the hyperfine spin and $m = -f,-f + 1,...,f$ is the magnetic sublevel. The wave function is normalized as $\int d\bm{r} |\sum_{m}\psi _m^{(f)}(\bm{r})|^2 = N_f$, where $N_f$ is the number of spin-$f$ atoms. The total energy of the mixture of spin-1 and spin-2 BECs is written as
\begin{equation}
	E = E^{(1)} + E^{(2)} + E^{(12)}.
	\label{eq:totalE}
\end{equation}
Here and henceforth, superscripts $(1)$, $(2)$, and $(12)$ refer to the hyperfine spins. The energy of each spin component is given by
\begin{equation}
	\begin{aligned}[b]
	E^{(1)} = & \int {\rm d}\bm{r} \sum\limits_{m =  - 1}^1 {\psi_m^{(1)*}(\bm{r}) \left[-{\frac{{{\hbar ^2}}}{{2M}}{\nabla ^2} + V_1(\bm{r})} \right]\psi _m^{(1)}}(\bm{r}) \\
	 &+\frac{1}{2}\int {\rm d}\bm{r} \Big[ g_0^{(1)} + g_1^{(1)}{\bm{F}^{(1)}}(\bm{r})\cdot{\bm{F}^{(1)}}(\bm{r}) \Big]\rho_1^2(\bm{r}),
	\end{aligned}
	\label{eq:e1}
\end{equation}
\begin{equation}
	\begin{aligned}[b]
	E^{(2)} = & {\int {{\rm d}\bm{r}} \sum\limits_{m =  - 2}^2 {\psi _m^{(2)*}(\bm{r})} \left[ { - \frac{{{\hbar ^2}}}{{2M}}{\nabla ^2} + {V_2}(\bm{r})} \right]\psi _m^{(2)}(\bm{r})} \\
	& + \frac{1}{2} \int {{\rm d}\bm{r}} {\left[ g_0^{(2)} + g_1^{(2)}{\bm{F}^{(2)}}(\bm{r})\cdot{\bm{F}^{(2)}}(\bm{r} ) \right.}\\ 
	& {\left.  +\ g_2^{(2)}{{\left| {A_0^{(2)}(\bm{r})} \right|}^2} \right]} \rho _2^2(\bm{r}),
	\end{aligned}
	\label{eq:e2}
\end{equation}
where $M$ is the mass of an atom and $V_1(\bm{r})$ and $V_2(\bm{r})$ are the external potentials. The interaction coefficients in Eqs.~(\ref{eq:e1}) and (\ref{eq:e2}) are defined as
\begin{subequations}
	\begin{align}
		g_0^{(1)} &= \frac{{4\pi {\hbar ^2}}}{M}\frac{{a_0^{(1)} + 2a_2^{(1)}}}{3},\\
		g_1^{(1)} &= \frac{{4\pi {\hbar ^2}}}{M}\frac{{a_2^{(1)} - a_0^{(1)}}}{3},\\
		g_0^{(2)} &= \frac{{4\pi {\hbar ^2}}}{M}\frac{{4a_2^{(2)} + 3a_4^{(2)}}}{7},\\
		g_1^{(2)} &= \frac{{4\pi {\hbar ^2}}}{M}\frac{{a_4^{(2)} - a_2^{(2)}}}{7},\\
		g_2^{(2)} &= \frac{{4\pi {\hbar ^2}}}{M}\frac{{7a_0^{(2)} - 10a_2^{(2)} + 3a_4^{(2)}}}{7},
	\end{align}
\end{subequations}
where $a_\mathcal{F}^{(f)}$ is the $s$-wave scattering length of a
collision channel with the total spin $\mathcal{F}$. In general, the
macroscopic wave function can be decomposed into $\psi_m^{(f)}(\bm{r})
= \sqrt{\rho_f(\bm{r})} \zeta _m^{(f)}(\bm{r})$, where
$\rho_f(\bm{r})$ is the density of spin-$f$ atoms and $\zeta
_m^{(f)}(\bm{r})$ is the spin wave function satisfying $\sum_{m}
|\zeta _m^{(f)}(\bm{r})|^2 = 1$. Using the spin wave function $\zeta
_m^{(f)}(\bm{r})$, we define the magnetization vector fields in Eqs.~(\ref{eq:e1}) and (\ref{eq:e2}) as
\begin{equation}
	\begin{aligned}
		{\bm{F}^{(f)}}(\bm{r}) = \sum\limits_{mm'} {\zeta _m^{(f)*}(\bm{r})\bm{f}_{mm'}^{(f)}} \zeta _{m'}^{(f)}(\bm{r}),
	\end{aligned}
	\label{eq:two}
\end{equation}
where $\bm{f}^{(f)}$ is the vector of spin-$f$ matrices. The spin-singlet scalar for spin-2 in Eq.~(\ref{eq:e2}) is defined as
\begin{equation}
A_0^{(2)} = \frac{1}{{\sqrt 5 }}\left( {2\zeta _2^{(2)}\zeta _{ - 2}^{(2)} - 2\zeta _1^{(2)}\zeta _{ - 1}^{(2)} + \zeta {{_0^{(2)}}}\zeta {{_0^{(2)}}}} \right).
\label{eq:spinsinglet}
\end{equation}
The interaction energy between spin-1 and spin-2 components is given by \cite{irikura2018}
\begin{equation}\label{eq:e12}
	\begin{aligned}[b]
		E^{(12)} = & \int {{\rm d}\bm{r}} {\left[ g_0^{(12)} + g_1^{(12)}{\bm{F}^{(1)}}(\bm{r})\cdot{\bm{F}^{(2)}}(\bm{r}) \right.}\\
		&{\left. + g_2^{(12)}P_1^{(12)}(\bm{r}) \right]}{\rho _1}(\bm{r}){\rho _2}(\bm{r}),
	\end{aligned}
\end{equation}
where the interaction coefficients have the forms
\begin{subequations}
	\begin{align}
		g_0^{(12)} &= \frac{{4\pi {\hbar ^2}}}{M}\frac{{2a_2^{(12)} + a_3^{(12)}}}{3},\\
		g_1^{(12)} &= \frac{{4\pi {\hbar ^2}}}{M}\frac{{a_3^{(12)} - a_2^{(12)}}}{3},\\
		g_2^{(12)} &= \frac{{4\pi {\hbar ^2}}}{M}\frac{{3a_1^{(12)} - 5a_2^{(12)} + 2a_3^{(12)}}}{3}.
	\end{align}
\end{subequations}
In Eq.~(\ref{eq:e12}), we defined
\begin{equation}
	P_1^{(12)} = {\left| {{A_{1,1}}} \right|^2} + {\left| {{A_{1,0}}} \right|^2} + {\left| {{A_{1, - 1}}} \right|^2},
\end{equation}
where
\begin{subequations}
	\begin{align}
		&A_{1,1} =\frac{1}{\sqrt{10}}\zeta _1^{(1)}\zeta _0^{(2)} - \sqrt {\frac{3}{{10}}} \zeta _0^{(1)}\zeta _1^{(2)} + \sqrt {\frac{3}{5}} \zeta _{ - 1}^{(1)}\zeta _2^{(2)},\\
		&A_{1,0} = \sqrt {\frac{3}{{10}}} \zeta _1^{(1)}\zeta _{ - 1}^{(2)} - \sqrt {\frac{2}{5}} \zeta _0^{(1)}\zeta _0^{(2)} + \sqrt {\frac{3}{{10}}} \zeta _{ - 1}^{(1)}\zeta _1^{(2)},\\
		&A_{1, - 1} = \sqrt {\frac{3}{5}} \zeta _1^{(1)}\zeta _{ - 2}^{(2)} - \sqrt {\frac{3}{{10}}} \zeta _0^{(1)}\zeta _{ - 1}^{(2)} +  \frac{1}{\sqrt{10}} \zeta _{ - 1}^{(1)}\zeta _0^{(2)}.
	\end{align}
\end{subequations}
In the present study, we neglect the effects of the external magnetic field and the magnetic dipole-dipole interaction.

The coupled Gross-Pitaevskii (GP) equations are obtained by the functional derivative of the total energy as
\begin{equation}
	\begin{aligned}
		i\hbar \frac{{\partial \psi _m^{(f)}}}{{\partial t}} = \frac{{\delta E}}{{\delta\psi _m^{(f)*}}}.
	\end{aligned}
	\label{eq:gpes}
\end{equation}
To obtain the ground state, we propagate the GP equation in imaginary time, where $i$ on the left-hand side of Eq.~(\ref{eq:gpes}) is replaced by $-1$. The real- and imaginary-time evolutions are numerically integrated using the fourth-order Runge-Kutta method with the pseudospectral scheme.

\begin{figure}[bt]
	\includegraphics[scale=0.5]{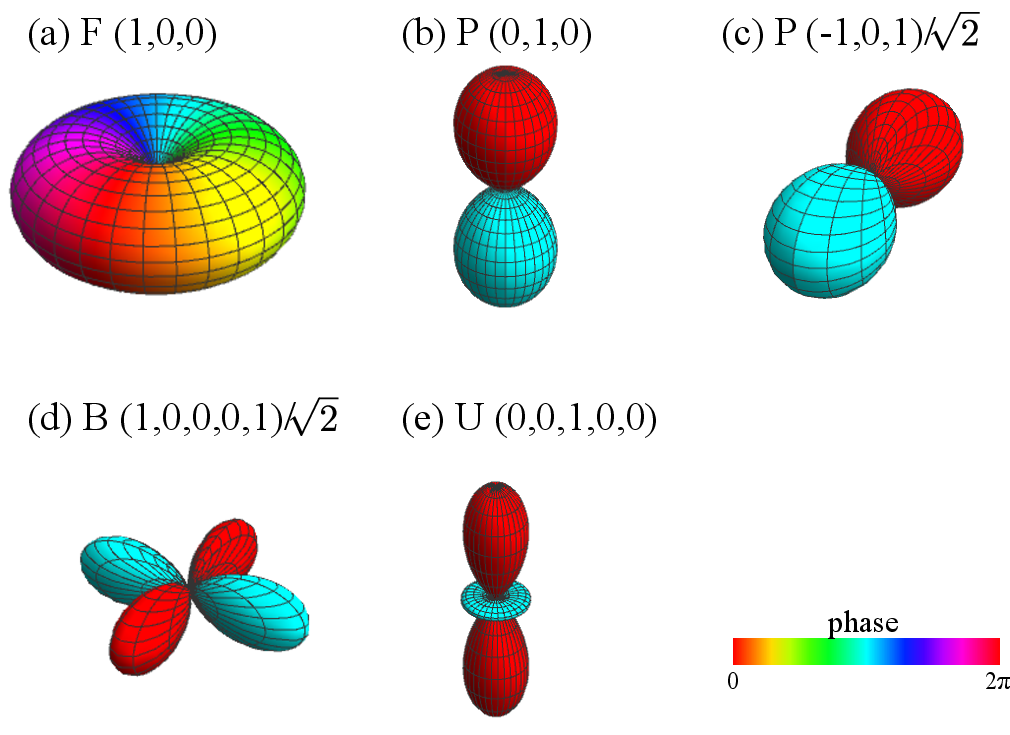} 
	\caption{Spherical harmonic representations of the spin-1 and
          spin-2 states. The surface and its color represent the
          isosurface and phase of the complex function $S\left(
          {\theta ,\phi } \right)$. (a) Ferromagnetic state
          (abbreviated by F) and (b, c) polar state (P) of spin-1. (d) Biaxial nematic state (B) and (e) uniaxial nematic state (U) of spin-2. Spin vectors are denoted by $\bm{\zeta}^{(1)} = ({\zeta}_1^{(1)},{\zeta}_0^{(1)},{\zeta}_{-1}^{(1)})$ and $\bm{\zeta}^{(2)} = ({\zeta}_2^{(2)},{\zeta}_1^{(2)},{\zeta}_0^{(2)},{\zeta}_{-1}^{(2)},{\zeta}_{-2}^{(2)})$.}
	\label{fig:spherical}
\end{figure}

For visualizing the symmetry of spin states, it is convenient to use the spherical harmonic representation
\begin{equation} \label{eq:sphe}
	S\left( {\theta ,\phi } \right) = \sum\limits_{m = {-f}}^f {\zeta _m^{(f)}Y_f^m\left( {\theta ,\phi } \right)},
\end{equation}
where $Y_f^m\left( {\theta ,\phi } \right)$ is the spherical harmonics. Figure~\ref{fig:spherical} shows several examples of the spherical harmonic representations of spin states. Henceforth, we use the abbreviations, F, P, B, and U for the spin states (Fig.~\ref{fig:spherical}).

In the present paper, we restrict ourselves to the BECs of
${}^{87}{\rm{Rb}}$ atoms. The $s$-wave scattering lengths for spin-1
${}^{87}{\rm{Rb}}$ atoms are known to be $a_0^{(1)} = 101.8{a_B}$,
$a_2^{(1)} = 100.4a_B$ \cite{klausen2001}, where $a_B$ is the Bohr
radius. For these values, the ground state of the spin-1 BEC is the F
state (Fig.~\ref{fig:spherical}(a)). The scattering lengths for spin-2
atoms were measured to be $a_2^{(2)} - a_0^{(2)} = 3.51{a_B}$,
$a_4^{(2)} - a_{2}^{(2)} = 6.95{a_B}$~\cite{widera2006}, and $(4a_4^{(2)} + 3a_2^{(2)})/7 = 95.44a_B$~\cite{egorov2013}, which give $a_0^{(2)} = 87.96a_B$, $a_2^{(2)} = 91.47a_B$, and $a_4^{(2)} = 98.42a_B$.
For these values, the ground state of the spin-2 BEC is a linear combination of the B and U states (Figs.~\ref{fig:spherical}(d) and \ref{fig:spherical}(e)).
In Ref.~\cite{eto2018}, the scattering lengths between spin-1 and spin-2
were measured to be $a_3^{(12)} - a_2^{(12)} = 2.5 a_B$ and
$a_1^{(12)} - a_2^{(12)} = 3.1 a_B$. Combining these values with $(3
a_1^{(12)} + 5 a_2^{(12)} + 2 a_3^{(12)}) / 10 = 98.006 a_B$ reported
in Ref.~\cite{egorov2013}, we can determine all the interspin
scattering lengths as $a_1^{(12)} = 99.68a_B$, $a_2^{(12)} =
96.58a_B$, and $a_3^{(12)} = 99.08a_B$. On the other hand, the
experiment in Ref.~\cite{gomez2019} gave $a_3^{(12)} - a_2^{(12)}=1.36
a_B$ and $a_1^{(12)} - a_2^{(12)}=1.40 a_B$. The corresponding
interspin scattering lengths are $a_1^{(12)} = 98.71a_B$, $a_2^{(12)}
= 97.31a_B$, and $a_3^{(12)} = 98.67a_B$. We refer to these two sets of scattering lengths based on Refs.~\cite{eto2018} and \cite{gomez2019} as ``set I'' and ``set II'', respectively.

\section{Ground states}
\label{sec:third}
In this section, we present the ground states of a mixture of spin-1 and spin-2 ${}^{87}{\rm{Rb}}$ BECs. To obtain the ground state, the imaginary-time evolution is started from initial states with random complex numbers. We repeat this procedure numerous times to ensure that the obtained state is the true ground state. The global phase rotation and spin rotation are applied to the obtained state appropriately, since the system has the U(1) symmetry and spin-rotation symmetry.

\subsection{Single-mode approximation}\label{subsec:sma}
\begin{figure}[tb]
	\includegraphics[scale=0.77]{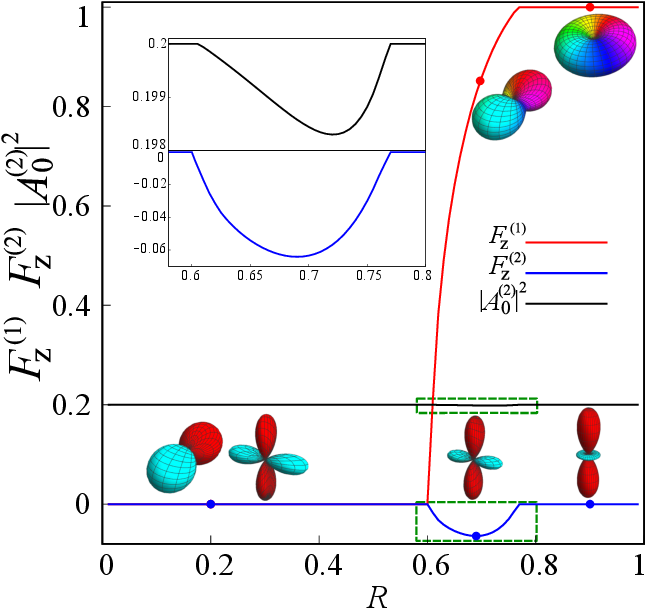} 
	\caption{Ground state under the single-mode approximation. The
		lines represent ${F_z^{(1)}}$, ${F_z^{(2)}}$, and
		$|A_0^{(2)}|^2$ as functions of the atomic number ratio $R =
		N_1 / (N_1 + N_2)$, where the state is rotated so that the
		magnetizations in the $x$-$y$ direction, $F_ \bot ^{(1)}$
		and $F_ \bot ^{(2)}$, vanish. The spherical-harmonic
		representations of spin-1 and spin-2 are shown for $R =
		0.2$, $0.7$, and $0.9$. The insets show magnifications of the dashed rectangle regions.}
	\label{fig:f_rho}
\end{figure}
First, we consider the ground state under the single-mode
approximation, which is valid if the size of the atomic cloud is much
smaller than the spin healing lengths. We assume that $\rho_1(\bm{r})$
and $\rho_2(\bm{r})$ are fixed at the same distribution and that
$\bm{\zeta}^{(f)}$ does not depend on the position. We define the
atomic number ratio as $R = N_1 / (N_1 + N_2)$. In
Ref.~\cite{irikura2018}, only the case of $R = 0.5$ was studied and
the ground state for ${}^{87}{\rm{Rb}}$ was shown to be the P state
for spin-1 and B state for spin-2 (we hereafter refer to this state as
``PB''). Here we extend this result to other values of $R$.

Figure \ref{fig:f_rho} shows the $R$ dependence of the ground
state. For $0 < R \lesssim 0.6$, the ground state is the PB state,
consistent with the results in Ref.~\cite{irikura2018}.  For $0.6 \lesssim R \lesssim 0.77$, both spin-1 and spin-2 components acquire magnetization with opposite directions. This state corresponds to the $a_-$ state defined in Ref.~\cite{irikura2018}. For $R \gtrsim 0.77$, the ground state becomes the F state for spin-1 and U state for spin-2. In the limit of $R \rightarrow 0$ and $R \rightarrow 1$, this result is consistent with the well-known ground state of an individual spin-1 or spin-2 BEC. In Fig~\ref{fig:f_rho}, parameter set I of the scattering lengths is used; we have confirmed that set II also gives qualitatively the same result.

\subsection{Phase separation}\label{subsec:1Ddepend}
\begin{figure}[bt]
	\includegraphics[scale=0.7]{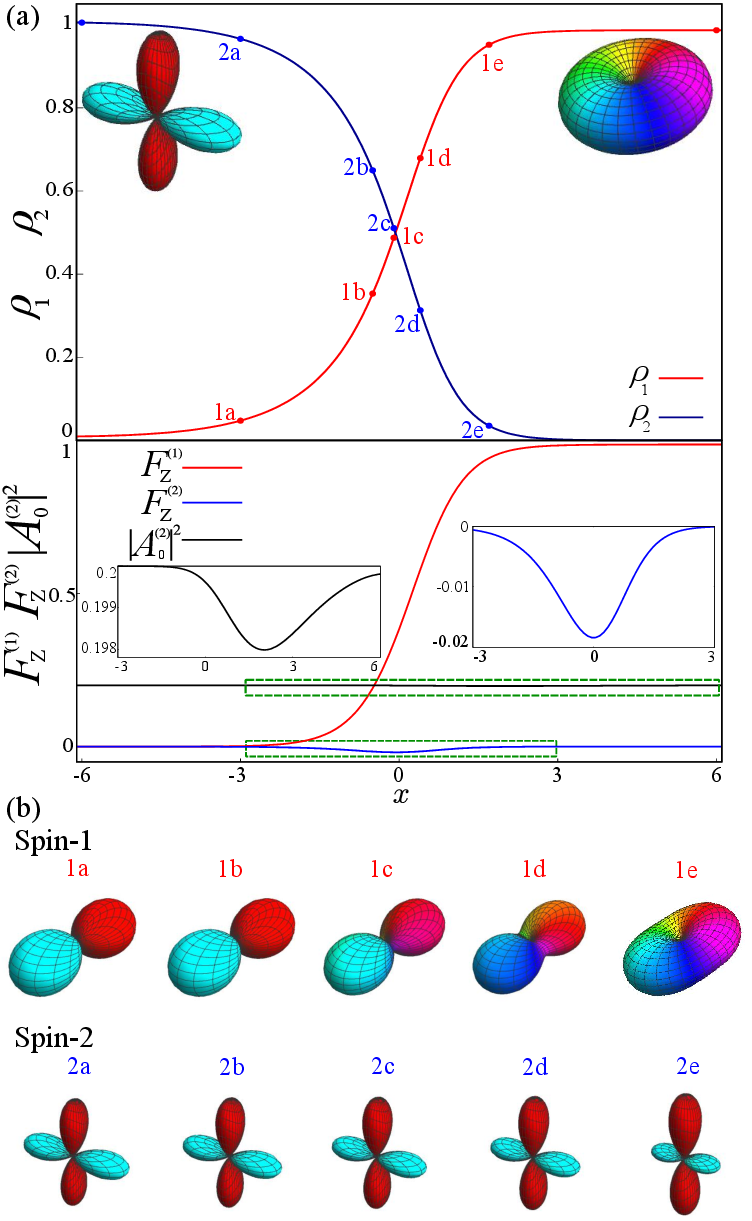} 
	\caption{Ground state of the one-dimensional system. (a)
          $\rho_1(x)$, $\rho_2(x)$ (upper panel), $F_z^{(1)}(x)$,
          $F_z^{(2)}(x)$, and $|A_0^{(2)}(x)|^2$ (lower panel). The
          magnetizations in the $x$-$y$ direction  are zero. The
          insets show magnifications of $|A_0^{(2)}(x)|^2$ and
          $F_z^{(2)}(x)$ in the dashed rectangle. (b)
          Spherical-harmonic representations of the spin-1 and spin-2 states at the positions marked in (a). Parameter set I is used.}
	\label{fig:gs1d}
\end{figure}	
To study the miscibility of the spin-1 and spin-2 BECs, we consider a one-dimensional system without external potentials, $V_1 = V_2 = 0$. We normalize the length and density by $L = 1/\sqrt{4\pi a_{B} n_0}$ and $n_0$, respectively, where $n_0$ is the average density of both components.

Figure \ref{fig:gs1d} shows the density and spin distributions of the ground state, which is the main result in the former part of the paper.
The ground state exhibits phase separation between spin-1 and spin-2.
In the limits of $x \rightarrow +\infty$ and $x \rightarrow -\infty$
(i.e., deep in the spin-1 and spin-2 sides), the spin state approaches
the F state of spin-1 and the nematic state of spin-2 (a linear
combination of the U and B states), respectively, consistent with the
results in Fig.~\ref{fig:f_rho}. The behavior near the interface is
also similar to that in the intermediate region of $R$ in Fig.~\ref{fig:f_rho}: spin-1 transforms between the P and F states, whereas spin-2 exhibits magnetization opposite to that of spin-1.
It is interesting to note that the spin-2 magnetization only emerges
near the interface, which is attributed to the interaction with the spin-1 component.
We have confirmed that both parameter sets I and II give qualitatively the same result.

\section{Uniformly mixed metastable state} \label{sec:fourth}
\subsection{Metastability analysis}\label{subsec:metaAna}
In this section, we show that the uniformly mixed PB state can be a metastable state depending on the scattering lengths, which is the main result in the latter part of this paper.
Before showing the metastability, we first confirm that the energy of the uniformly mixed PB state is larger than the separated ground state shown in Fig.~\ref{fig:gs1d}.
The total energy of the uniformly mixed PB state in a volume $V$ containing $N_1$ and $N_2$ atoms of spin-1 and spin-2 is given by

\begin{equation}\label{eq:E_PB}
	E_{{\rm{mix}}}^{{\rm{PB}}} = \frac{{N_1^2}}{{2V}}g_0^{(1)} + \frac{{N_2^2}}{{2V}}\left( {g_0^{(2)} + \frac{1}{5}g_2^{(2)}} \right) + \frac{{{N_1}{N_2}}}{V}g_0^{(12)}.
\end{equation}
As shown in Fig.~\ref{fig:gs1d}, the ground state exhibits phase
separation between the F state of spin-1 and the nematic state of
spin-2 (we refer to this state as ``FN''). For a sufficiently large system, the bulk energy of each separated region is dominant and the energy of the interface layer can be neglected.
In this case, the energy is evaluated to be
\begin{equation}
	E_{\rm{separate}}^{{\rm{FN}}} = \frac{{N_1^2}}{{2V_1}} \left( {g_0^{(1)} + g_1^{(1)}} \right) + \frac{{N_2^2}}{{2V_2}}\left( {g_0^{(2)} + \frac{1}{5}g_2^{(2)}} \right),
\end{equation}
where $V_1$ and $V_2$ are the volume of each separated region, satisfying $V_1 + V_2 = V$.
The values of $V_1$ and $V_2$ are determined such that the pressures of two regions are balanced as
\begin{equation}\label{eq:pressure}
	\frac{N_1^2}{2 V_1^2} \left(g_0^{(1)} + g_1^{(1)}\right) = \frac{N_2^2}{2 V_2^2} \left(g_0^{(2)} + \frac{1}{5} g_2^{(2)}\right).
\end{equation}
Using Eqs.~(\ref{eq:E_PB})-(\ref{eq:pressure}), we obtain the difference between the two energies
\begin{equation}
	\begin{aligned}[b]
		&E_{{\rm{mix}}}^{{\rm{PB}}} - E_{\rm{separate}}^{{\rm{FN}}} =  - \frac{{N_1^2}}{{2V}}g_1^{(1)} + \frac{N_{1}N_{2}}{V} \\
		&\times \Biggl\{g_0^{(12)} - {\left[{\left( {g_0^{(1)} + g_1^{(1)}} \right)\left( {g_0^{(2)} + \frac{1}{5}g_2^{(2)}} \right)}\right]}^{1/2} \Biggr\}, 
	\end{aligned}
\end{equation}
which is always positive for both parameter sets I and II.

We next examine the stability of the uniformly mixed PB state using
the Bogoliubov analysis. We divide the wave function into the
uniformly mixed state $\bm{\Psi}_{\rm PB}$ and a small deviation $\bm{\phi}(\bm{r}, t)$ as
\begin{equation}\label{eq:Bogo}
	 {\bm{\Psi}}^{(f)}({\bm{r}},t) = e^{-i{\mu^{(f)}t/\hbar}}\Big[{\bm{\Psi}}^{(f)}_{\rm {PB}}+\bm{\phi}^{(f)}(\bm{r},t)\Big],
\end{equation}
where the chemical potential of the P state of spin-1 is $\mu^{(1)} = g_0^{(1)} \rho_1 + g_0^{(12)} \rho_2$ and that of the B state of spin-2 is $\mu^{(2)} = (g_0^{(2)} + g_2^{(2)} / 5) \rho_2 + g_0^{(12)} \rho_1$. The small deviation $\bm{\phi}(r,t)$ is expanded as
\begin{equation}\label{eq:phi}
	\bm{\phi}(\bm{r},t) = \sum_{\bm{k}} \left[\bm{u}(\bm{k})e^{i(\bm{k\cdot r}-\omega_kt)} + \bm{v}^{*}(\bm{k})e^{-i(\bm{k \cdot r}-\omega_k^*t)} \right] .	
\end{equation}
Substituting Eqs.~(\ref{eq:Bogo}) and (\ref{eq:phi}) into the GP equations in Eq.~(\ref{eq:gpes}) and neglecting the second- and third-order terms of $\bm{u}(\bm{k})$ and $\bm{v}(\bm{k})$, we obtain the Bogoliubov-de Gennes equation.
The eigenenergies $\hbar\omega_k$ of the Bogoliubov-de Gennes equation are given in \ref{appendix:bogo}. For the stable system, all the eigenenergies must be real and positive. If eigenenergies for some wave number $\bm{k}$ are complex, the corresponding eigenmodes grow exponentially in time and the uniformly mixed PB state is dynamically unstable.
For the interaction coefficients in set I, all the eigenenergies are found to be real and positive for $R = N_1 / (N_1 + N_2) \lesssim 0.6$, which indicates that the uniformly mixed PB state is metastable for $R \lesssim 0.6$.
However, applying the interaction coefficients in set II, we find that complex eigenenergies appear for any $R$; therefore, the uniformly mixed PB state is dynamically unstable against phase separation for set II.

\subsection{Phase separation dynamics via nucleation}
To confirm the metastability of the uniformly mixed PB state, we numerically solve the real-time evolution of the one-dimensional GP equation for parameter set I starting from the uniformly mixed PB state.
To trigger the phase separation, we add the term
\begin{equation}\label{eq:localPer}
	B_0 e^{-x^2 / a^2} g_f \sum_{m'} (f_x)_{mm'}^{(f)} \psi_{m'}(x, t)
\end{equation}
to the right-hand side of the GP equation, where $g_1 = -1/2$ and $g_2 = 1/2$.
Such a synthetic local magnetic field can be generated by a laser beam \cite{Grimm2000, Chiba2008, Kim2020}.

\begin{figure}[bt]
	\includegraphics[scale=0.665]{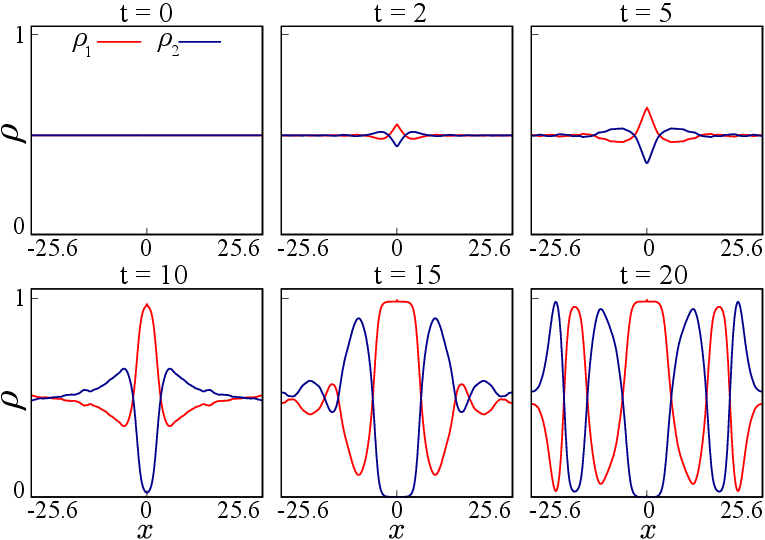} 
	\caption{Dynamics of the one-dimensional system starting from the uniformly mixed PB state with the mixing ratio $R = 0.5$, where the local perturbation in Eq.~(\ref{eq:localPer}) is applied with $a = 0.1$ and $B_0 = 2$.
		Red (light-gray) and blue (dark-gray) lines show
                spin-1 and spin-2 density distributions,
                respectively. Length and time are normalized by $L = 1
                / \sqrt{4\pi a_B n_0}$ and $M L^2 / \hbar$,
                respectively, where $n_0$ is the average density. The parameter set I is used. See the Supplemental Material for a movie of the dynamics \cite{SM}.}
	\label{fig:PB}
\end{figure}

Figure~\ref{fig:PB} shows the time evolution of the density distributions of spin-1 and spin-2 components for $a = 0.1$ and $B_0 = 2$. The local perturbation at the center triggers the phase separation, which extends over the whole space.
We confirmed that the phase separation does not occur for $B_0 \lesssim 0.1$, which indicates that the uniformly mixed PB state is metastable.

\subsection{Simple explanation of the metastability}
Here, we provide a simple explanation for why the uniformly mixed PB state is metastable, i.e., why there is an energy barrier against phase separation.
Let us consider the uniformly mixed PB state with a ratio $R$, say $R = 0.5$.
Suppose that the phase separation begins as $R(\bm{r}) =\rho_1(\bm{r}) / [\rho_1(\bm{r}) + \rho_2(\bm{r})]= 0.5 + \epsilon(\bm{r})$, where $\epsilon(\bm{r}) \ll 1$.
According to Fig.~\ref{fig:f_rho} under the SMA, the lowest-energy
spin state is the PB state around $R = 0.5$; therefore, the local spin
state is fixed to the PB state even when the modulation
$\epsilon(\bm{r})$ is present.
In this case, the interaction energy is given by
\begin{equation}\label{eq:energyTerm}
	\int d\bm{r} \left[ \frac{1}{2} g_0^{(1)} \rho_1^2 + \frac{1}{2} \left( g_0^{(2)} + \frac{1}{5} g_2^{(2)} \right) \rho_2^2 + g_0^{(12)} \rho_1 \rho_2 \right].
\end{equation}
Hence, spin-1 and spin-2 are miscible (immiscible) for $g_0^{(1)} \left(g_0^{(2)} + g_2^{(2)} / 5\right) - \left(g_0^{(12)}\right)^2 > 0$ ($< 0$).
The miscible (immiscible) condition is satisfied for the parameter set I (set II).
Thus, for the parameter set I, the energy is increased by a small modulation $\epsilon(\bm{r})$, which makes the uniformly mixed PB state metastable.
When $R(\bm{r})$ deviates substantially from 0.5, the local spin state no longer remains in the PB state and Eq.~(\ref{eq:energyTerm}) cannot be used. As a result, the phase separation can reduce the energy, which results in the dynamics shown in Fig.~\ref{fig:PB}.

\section{Experimental proposal}
\label{sec:fifth}
\begin{figure*}[tb]
	\includegraphics[scale=0.65]{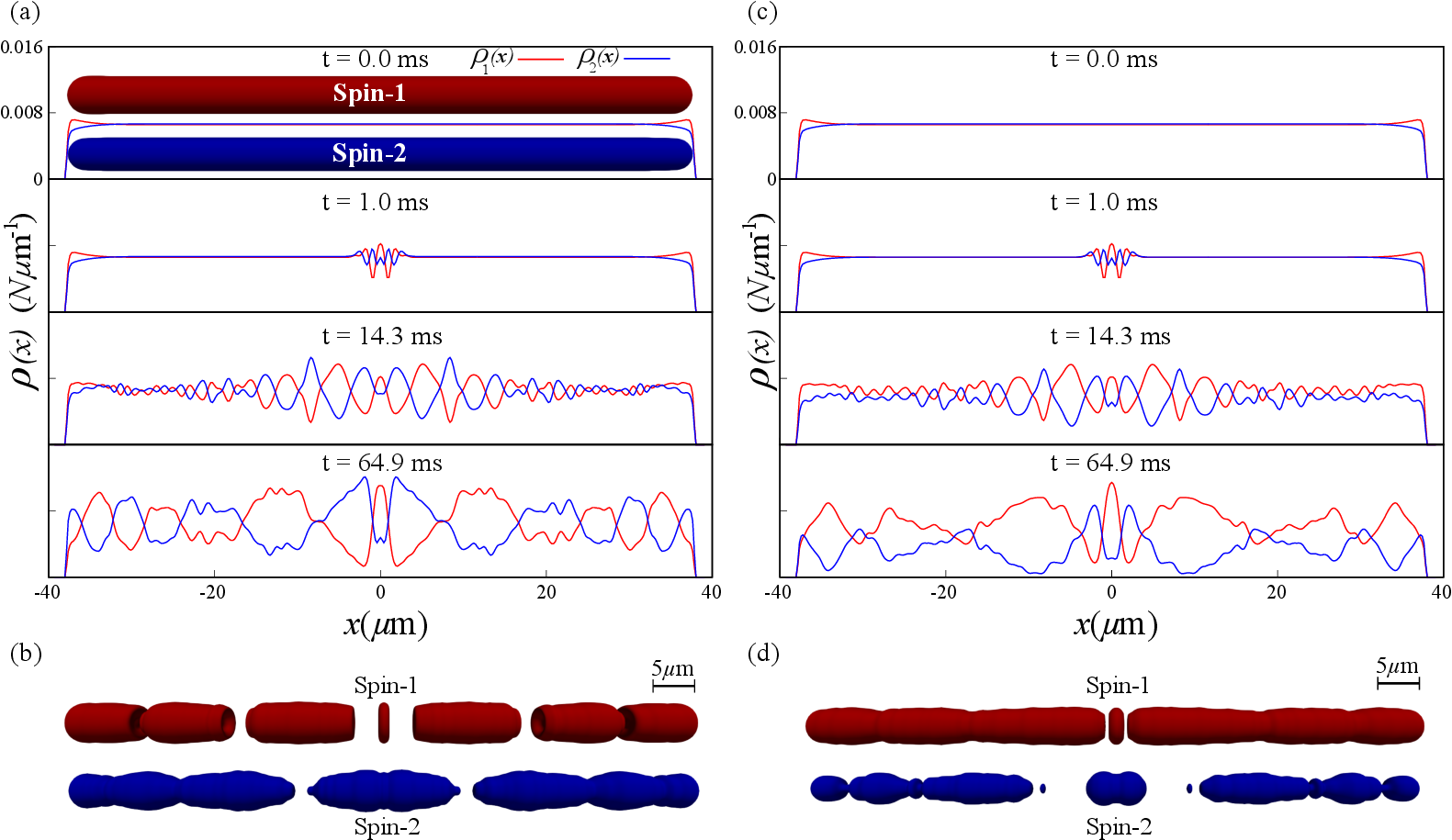} 
	\caption{(a), (c) Time evolution of the one-dimensional
          density distributions $\rho_f(x) = \int \rho_f(\bm{r}, t) dy
          dz$ and (b), (d) isodensity surfaces at $t \simeq 65$ ms for
          the system trapped in the potential given in
          Eq.~(\ref{eq:trap}). The initial state is the metastable PB
          state, and the local perturbation in Eq.~(\ref{eq:localPer})
          is added at $t=0\ \text{ms}$. The atomic loss is not
          included in (a) and (b), and included in (c) and (d). The parameter set I is used. See the Supplemental Material for movies of the dynamics \cite{SM}.}
	\label{fig:dynamic}
\end{figure*}
We consider a realistic three-dimensional system confined in a radially harmonic and axially box-like potential as
\begin{equation}\label{eq:trap}
	V_1 = V_2 = \frac{M \omega_\perp^2}{2} (y^2 + z^2) + V_0 \theta(x - x_0) \theta(-x - x_0),
\end{equation}
where $\omega_\perp = 2\pi \times 250$ Hz, $x_0 = 38$ $\mu{\rm m}$, $\theta$ is the Heaviside step function, and $V_0$ is taken to be much larger than the chemical potential.
The box-like potential in the $x$ direction avoids complexity arising from inhomogeneous density distribution for, e.g., a weak harmonic potential.
The number of $^{87}{\rm Rb}$ atoms is $N_1 = N_2 = 2 \times 10^5$.

We prepare the initial state as follows. First, the ground state $\psi_g(\bm{r})$ of the $|f = 1, m = -1\rangle$ state for $N_1 = 4 \times 10^5$ is prepared using the imaginary-time evolution of the GP equation.
This wave function is then transferred to the PB state as
$\bm{\psi}^{(1)}(\bm{r}) = \psi_g(\bm{r}) (0, 1, 0) / \sqrt{2}$ and
$\bm{\psi}^{(2)}(\bm{r}) = \psi_g(\bm{r}) (1, 0, 0, 0, 1) / 2$, which
is experimentally possible using microwave and radio-frequency
fields. The obtained PB state has a cigar shape of length $\simeq 76$
$\mu{\rm m}$, as shown in Fig.~\ref{fig:dynamic}(a) ($t = 0$ ms).

In the real-time evolution, the local perturbation in Eq.~(\ref{eq:localPer}) is added, where $B_0 = 4 \hbar \omega_\perp$ and $a = 1\,\mu{\rm m}$.
Figure~\ref{fig:dynamic}(a) shows the dynamics of the density distributions.
The local perturbation around $x = 0$ triggers the spin modulation, and the phase separation spreads over the whole space, as shown in Figs.~\ref{fig:dynamic}(a) and \ref{fig:dynamic}(b).

The spin-2 atom has a higher energy than the spin-1 atom as a result
of the hyperfine splitting.
If the transition from spin-2 to spin-1 occurs in collisional processes, the relevant atoms escape from the system.
This effect can be taken into account in the GP equation by adding imaginary parts to the interaction coefficients, which makes the time evolution nonunitary to simulate the atomic loss~\cite{tojo2009spin}.
For the inelastic two-body collisions between spin-2 atoms, we replace the interaction coefficients as
\begin{subequations}
	\begin{align}
		\bar g_0^{(2)} &= g_0^{(2)} - \frac{2}{7} i\hbar{b_2}\\
		\bar g_1^{(2)} &= g_1^{(2)} + \frac{1}{14} i\hbar{b_2}\\
		\bar g_2^{(2)} &= g_2^{(2)} + \frac{{5}}{7} i\hbar{b_2} - \frac{1}{2} i\hbar{b_0}
	\end{align}
\end{subequations}
where ${b_0} = 9.9 \times {10^{ -
    14}}\,{\rm{c}}{{\rm{m}}^{\rm{3}}}{\rm{/s}}$ and ${b_2} = 24.3
\times {10^{ - 14}}\,{\rm{c}}{{\rm{m}}^{\rm{3}}}{\rm{/s}}$ are the
loss coefficients for collision channels with total spins 0 and 2, respectively~\cite{tojo2009spin}. We ignore the two-body inelastic loss due to collisions between spin-1 and spin-2 atoms, since the relevant loss coefficients are much smaller than $b_0$ and $b_2$ \cite{Eto2016, ref1}.
Figures~\ref{fig:dynamic}(c) and \ref{fig:dynamic}(d) show the results with the atomic loss.
Although the spin-2 atoms decrease in time, the patterns of the
density distributions are similar to those of the system without atomic loss.
Thus, the phase separation via nucleation can be observed in a realistic experimental system.

\section{Conclusions}\label{sec:concl}
We investigated the mixture of spin-1 and spin-2 ${}^{87}{\rm{Rb}}$
BECs, including the spatial degree of freedom within the mean-field
approximation. We showed that the ground state exhibits phase
separation between the spin-1 ferromagnetic state and the spin-2
nematic state. In the interface region between them, another phase
appears in which both spin-1 and spin-2 components have magnetizations
with opposite directions. We also found that the uniformly mixed state
of the spin-1 polar state and the spin-2 biaxial nematic state is
metastable for the $s$-wave scattering lengths measured in
Ref.~\cite{eto2018}. If we impart a local perturbation to this state,
phase separation via nucleation occurs. This phenomenon can be
observed in a realistic experiment even when the atomic loss due to
inelastic collisions occurs.

\begin{acknowledgments}
We thank Yujiro Eto for discussion of the inelastic loss rate between spin-1 and spin-2 $^{87}{\rm Rb}$ atoms. This work was supported by JSPS KAKENHI Grant Number JP23K03276.
\end{acknowledgments}

\appendix
\renewcommand{\thesubsection}{Appendix}
\subsection{BOGOLIUBOV ANALYSIS OF THE UNIFORMLY MIXED PB STATE}\label{appendix:bogo}
Diagonalizing the Bogoliubov-de Gennes equation derived from Eqs.~(\ref{eq:gpes}), (\ref{eq:Bogo}), and (\ref{eq:phi}), we obtain eight eigenvalues as
\begin{equation} \tag{A}
\begin{aligned}
\sqrt{ {\varepsilon _k}\left[ \varepsilon _k + (1-R) \left(8c_1^{(2)} -2c_2^{(2)}/5\right) \right] },\\
\sqrt{\left({\varepsilon _k} + 2Rc_2^{(12)}/5 \right) \left\{{\varepsilon _k} + 2/5 \left[Rc_2^{(12)}-(1-R) c_2^{(2)}\right] \right\}  }, \\
\sqrt{ \varepsilon_{k} \left( {\varepsilon _k} + Rc_0^{(1)} + (1-R)\left(c_0^{(2)} + c_2^{(2)}/5 \right) \pm \mathcal{M}^{1/2} \right)},\\
\sqrt{\mathcal{A} \pm \sqrt{\mathcal{B}}},
\label{eq:A}
\end{aligned}
\end{equation}
where $\varepsilon_k = (\hbar {\bf{k}} )^2/(2M)$, $c^{(f)}_n = n_0g^{(f)}_n$ with the total density $n_0$, and
\begin{widetext}
\begin{equation}\notag
\begin{aligned}
\mathcal{M}  & = 4R(1-R){{\left( {c_0^{(12)}} \right)}^2} + {{\left[ { Rc_0^{(1)} - (1-R)\left(c_0^{(2)} + c_2^{(2)}/5 \right)} \right]}^2}, \\
\mathcal{A}  & =  {\varepsilon _k}^2 + {\varepsilon _k} \left[ Rc_1^{(1)} + (1-R)\left(c_1^{(2)} - c_2^{(2)}/5 \right) + 3c_2^{(12)}/10 \right] \\
& + 3c_2^{(12)}/40 \left[ 4R(1-R) \left( c_1^{(1)} + c_1^{(2)} - c_2^{(2)}/5 - 2c_1^{(12)} \right) + 3\left(c_2^{(12)}\right)^2/5 \right], \\
\mathcal{B} & =  \alpha{\varepsilon _k}^2 + \beta{\varepsilon _k}  + \gamma
\end{aligned}
\end{equation}
with
\begin{equation}\notag
	\begin{aligned}
\alpha & = 1/4 \left[ 2Rc_1^{(1)} -2(1-R)\left( c_1^{(2)} - c_2^{(2)}/5 \right) - 3(2R-1)c_2^{(12)}/5 \right]^2 + R(1-R)\left(2c_1^{(12)} - 3c_2^{(12)}/5 \right)^2,\\
\beta &= 3c_2^{(12)}/40 \Biggl\{ 9\left(c_2^{(12)}\right)^2/25 + 6c_2^{(12)}/5 \left[R(3-4R)c_1^{(1)} - (1-R)\left((1-4R)\left(c_1^{(2)} - c_2^{(2)}/5\right) + 8Rc_1^{(12)} \right) \right]\\
& + 8R(1-R)\left[2\left(c_1^{(12)}\right)^2 + Rc_1^{(1)}\left(c_1^{(1)}-2c_1^{(12)}\right)+(1-R)\left(c_1^{(2)} - c_2^{(2)}/5\right)\left(c_1^{(2)} - c_2^{(2)}/5 - 2c_1^{(12)}\right) - \left(c_1^{(2)} - c_2^{(2)}/5 \right)c_1^{(1)} \right] \Biggr\},\\
\gamma &= 9\left(c_2^{(12)}\right)^2/1600  \left[ 4R(1-R)\left( c_1^{(1)} + c_1^{(2)} - c_2^{(2)}/5 - 2c_1^{(12)} \right) + 3c_2^{(12)}/5 \right]^2.\\
\end{aligned}
\end{equation}
The two eigenvalues $\sqrt{\mathcal{A} \pm \sqrt{\mathcal{B}}}$ in Eq.~(\ref{eq:A}) are doubly repeated eigenvalues.
\end{widetext}


\begin{thebibliography}{59}%
	\makeatletter
	\providecommand \@ifxundefined [1]{%
		\@ifx{#1\undefined}
	}%
	\providecommand \@ifnum [1]{%
		\ifnum #1\expandafter \@firstoftwo
		\else \expandafter \@secondoftwo
		\fi
	}%
	\providecommand \@ifx [1]{%
		\ifx #1\expandafter \@firstoftwo
		\else \expandafter \@secondoftwo
		\fi
	}%
	\providecommand \natexlab [1]{#1}%
	\providecommand \enquote  [1]{``#1''}%
	\providecommand \bibnamefont  [1]{#1}%
	\providecommand \bibfnamefont [1]{#1}%
	\providecommand \citenamefont [1]{#1}%
	\providecommand \href@noop [0]{\@secondoftwo}%
	\providecommand \href [0]{\begingroup \@sanitize@url \@href}%
	\providecommand \@href[1]{\@@startlink{#1}\@@href}%
	\providecommand \@@href[1]{\endgroup#1\@@endlink}%
	\providecommand \@sanitize@url [0]{\catcode `\\12\catcode `\$12\catcode
		`\&12\catcode `\#12\catcode `\^12\catcode `\_12\catcode `\%12\relax}%
	\providecommand \@@startlink[1]{}%
	\providecommand \@@endlink[0]{}%
	\providecommand \url  [0]{\begingroup\@sanitize@url \@url }%
	\providecommand \@url [1]{\endgroup\@href {#1}{\urlprefix }}%
	\providecommand \urlprefix  [0]{URL }%
	\providecommand \Eprint [0]{\href }%
	\providecommand \doibase [0]{https://doi.org/}%
	\providecommand \selectlanguage [0]{\@gobble}%
	\providecommand \bibinfo  [0]{\@secondoftwo}%
	\providecommand \bibfield  [0]{\@secondoftwo}%
	\providecommand \translation [1]{[#1]}%
	\providecommand \BibitemOpen [0]{}%
	\providecommand \bibitemStop [0]{}%
	\providecommand \bibitemNoStop [0]{.\EOS\space}%
	\providecommand \EOS [0]{\spacefactor3000\relax}%
	\providecommand \BibitemShut  [1]{\csname bibitem#1\endcsname}%
	\let\auto@bib@innerbib\@empty
	\bibitem [{\citenamefont {Vollhardt}\ and\ \citenamefont
		{Wolfle}(2013)}]{vollhardt2013superfluid}%
	\BibitemOpen
	\bibfield  {author} {\bibinfo {author} {\bibfnamefont {D.}~\bibnamefont
			{Vollhardt}}\ and\ \bibinfo {author} {\bibfnamefont {P.}~\bibnamefont
			{Wolfle}},\ }\href@noop {} {\emph {\bibinfo {title} {{The superfluid phases
					of helium 3}}}}\ (\bibinfo  {publisher} {Courier Corporation},\ \bibinfo
	{year} {2013})\BibitemShut {NoStop}%
	\bibitem [{\citenamefont {Norman}(2011)}]{norman2011challenge}%
	\BibitemOpen
	\bibfield  {author} {\bibinfo {author} {\bibfnamefont {M.~R.}\ \bibnamefont
			{Norman}},\ }\bibfield  {title} {\bibinfo {title} {{The challenge of
				unconventional superconductivity}},\ }\href
	{https://doi.org/10.1126/science.1200181} {\bibfield  {journal} {\bibinfo
			{journal} {Science}\ }\textbf {\bibinfo {volume} {332}},\ \bibinfo {pages}
		{196} (\bibinfo {year} {2011})}\BibitemShut {NoStop}%
	\bibitem [{\citenamefont {Ohmi}\ and\ \citenamefont
		{Machida}(1998)}]{Ohmi1998}%
	\BibitemOpen
	\bibfield  {author} {\bibinfo {author} {\bibfnamefont {T.}~\bibnamefont
			{Ohmi}}\ and\ \bibinfo {author} {\bibfnamefont {K.}~\bibnamefont {Machida}},\
	}\bibfield  {title} {\bibinfo {title} {{Bose-Einstein condensation with
				internal degrees of freedom in alkali atom gases}},\ }\href
	{https://doi.org/10.1143/JPSJ.67.1822} {\bibfield  {journal} {\bibinfo
			{journal} {J. Phys. Soc. Jpn.}\ }\textbf {\bibinfo {volume} {67}},\ \bibinfo
		{pages} {1822} (\bibinfo {year} {1998})}\BibitemShut {NoStop}%
	\bibitem [{\citenamefont {Ho}(1998)}]{Ho1998}%
	\BibitemOpen
	\bibfield  {author} {\bibinfo {author} {\bibfnamefont {T.-L.}\ \bibnamefont
			{Ho}},\ }\bibfield  {title} {\bibinfo {title} {{Spinor Bose condensates in
				optical traps}},\ }\href {https://doi.org/10.1103/PhysRevLett.81.742}
	{\bibfield  {journal} {\bibinfo  {journal} {Phys. Rev. Lett.}\ }\textbf
		{\bibinfo {volume} {81}},\ \bibinfo {pages} {742} (\bibinfo {year}
		{1998})}\BibitemShut {NoStop}%
	\bibitem [{\citenamefont {Law}\ \emph {et~al.}(1998)\citenamefont {Law},
		\citenamefont {Pu},\ and\ \citenamefont {Bigelow}}]{Law1998}%
	\BibitemOpen
	\bibfield  {author} {\bibinfo {author} {\bibfnamefont {C.~K.}\ \bibnamefont
			{Law}}, \bibinfo {author} {\bibfnamefont {H.}~\bibnamefont {Pu}},\ and\
		\bibinfo {author} {\bibfnamefont {N.~P.}\ \bibnamefont {Bigelow}},\
	}\bibfield  {title} {\bibinfo {title} {{Quantum spins mixing in spinor
				Bose-Einstein condensates}},\ }\href
	{https://doi.org/10.1103/PhysRevLett.81.5257} {\bibfield  {journal} {\bibinfo
			{journal} {Phys. Rev. Lett.}\ }\textbf {\bibinfo {volume} {81}},\ \bibinfo
		{pages} {5257} (\bibinfo {year} {1998})}\BibitemShut {NoStop}%
	\bibitem [{\citenamefont {Stamper-Kurn}\ \emph {et~al.}(1998)\citenamefont
		{Stamper-Kurn}, \citenamefont {Andrews}, \citenamefont {Chikkatur},
		\citenamefont {Inouye}, \citenamefont {Miesner}, \citenamefont {Stenger},\
		and\ \citenamefont {Ketterle}}]{Stamper1998}%
	\BibitemOpen
	\bibfield  {author} {\bibinfo {author} {\bibfnamefont {D.~M.}\ \bibnamefont
			{Stamper-Kurn}}, \bibinfo {author} {\bibfnamefont {M.~R.}\ \bibnamefont
			{Andrews}}, \bibinfo {author} {\bibfnamefont {A.~P.}\ \bibnamefont
			{Chikkatur}}, \bibinfo {author} {\bibfnamefont {S.}~\bibnamefont {Inouye}},
		\bibinfo {author} {\bibfnamefont {H.-J.}\ \bibnamefont {Miesner}}, \bibinfo
		{author} {\bibfnamefont {J.}~\bibnamefont {Stenger}},\ and\ \bibinfo {author}
		{\bibfnamefont {W.}~\bibnamefont {Ketterle}},\ }\bibfield  {title} {\bibinfo
		{title} {{Optical confinement of a Bose-Einstein condensate}},\ }\href
	{https://doi.org/10.1103/PhysRevLett.80.2027} {\bibfield  {journal} {\bibinfo
			{journal} {Phys. Rev. Lett.}\ }\textbf {\bibinfo {volume} {80}},\ \bibinfo
		{pages} {2027} (\bibinfo {year} {1998})}\BibitemShut {NoStop}%
	\bibitem [{\citenamefont {Stenger}\ \emph {et~al.}(1998)\citenamefont
		{Stenger}, \citenamefont {Inouye}, \citenamefont {Stamper-Kurn},
		\citenamefont {Miesner}, \citenamefont {Chikkatur},\ and\ \citenamefont
		{Ketterle}}]{stenger1998spin}%
	\BibitemOpen
	\bibfield  {author} {\bibinfo {author} {\bibfnamefont {J.}~\bibnamefont
			{Stenger}}, \bibinfo {author} {\bibfnamefont {S.}~\bibnamefont {Inouye}},
		\bibinfo {author} {\bibfnamefont {D.~M.}\ \bibnamefont {Stamper-Kurn}},
		\bibinfo {author} {\bibfnamefont {H.-J.}\ \bibnamefont {Miesner}}, \bibinfo
		{author} {\bibfnamefont {A.~P.}\ \bibnamefont {Chikkatur}},\ and\ \bibinfo
		{author} {\bibfnamefont {W.}~\bibnamefont {Ketterle}},\ }\bibfield  {title}
	{\bibinfo {title} {{Spin domains in ground-state Bose-Einstein
				condensates}},\ }\href {https://doi.org/10.1038/24567} {\bibfield  {journal}
		{\bibinfo  {journal} {Nature (London)}\ }\textbf {\bibinfo {volume} {396}},\
		\bibinfo {pages} {345} (\bibinfo {year} {1998})}\BibitemShut {NoStop}%
	\bibitem [{\citenamefont {Chang}\ \emph {et~al.}(2004)\citenamefont {Chang},
		\citenamefont {Hamley}, \citenamefont {Barrett}, \citenamefont {Sauer},
		\citenamefont {Fortier}, \citenamefont {Zhang}, \citenamefont {You},\ and\
		\citenamefont {Chapman}}]{Chang2004}%
	\BibitemOpen
	\bibfield  {author} {\bibinfo {author} {\bibfnamefont {M.-S.}\ \bibnamefont
			{Chang}}, \bibinfo {author} {\bibfnamefont {C.~D.}\ \bibnamefont {Hamley}},
		\bibinfo {author} {\bibfnamefont {M.~D.}\ \bibnamefont {Barrett}}, \bibinfo
		{author} {\bibfnamefont {J.~A.}\ \bibnamefont {Sauer}}, \bibinfo {author}
		{\bibfnamefont {K.~M.}\ \bibnamefont {Fortier}}, \bibinfo {author}
		{\bibfnamefont {W.}~\bibnamefont {Zhang}}, \bibinfo {author} {\bibfnamefont
			{L.}~\bibnamefont {You}},\ and\ \bibinfo {author} {\bibfnamefont {M.~S.}\
			\bibnamefont {Chapman}},\ }\bibfield  {title} {\bibinfo {title} {{Observation
				of spinor dynamics in optically trapped $^{87}\mathrm{Rb}$ Bose-Einstein
				condensates}},\ }\href {https://doi.org/10.1103/PhysRevLett.92.140403}
	{\bibfield  {journal} {\bibinfo  {journal} {Phys. Rev. Lett.}\ }\textbf
		{\bibinfo {volume} {92}},\ \bibinfo {pages} {140403} (\bibinfo {year}
		{2004})}\BibitemShut {NoStop}%
	\bibitem [{\citenamefont {Chang}\ \emph {et~al.}(2005)\citenamefont {Chang},
		\citenamefont {Qin}, \citenamefont {Zhang}, \citenamefont {You},\ and\
		\citenamefont {Chapman}}]{Chang2005}%
	\BibitemOpen
	\bibfield  {author} {\bibinfo {author} {\bibfnamefont {M.-S.}\ \bibnamefont
			{Chang}}, \bibinfo {author} {\bibfnamefont {Q.}~\bibnamefont {Qin}}, \bibinfo
		{author} {\bibfnamefont {W.}~\bibnamefont {Zhang}}, \bibinfo {author}
		{\bibfnamefont {L.}~\bibnamefont {You}},\ and\ \bibinfo {author}
		{\bibfnamefont {M.~S.}\ \bibnamefont {Chapman}},\ }\bibfield  {title}
	{\bibinfo {title} {{Coherent spinor dynamics in a spin-1 Bose condensate}},\
	}\href {https://doi.org/10.1038/nphys153} {\bibfield  {journal} {\bibinfo
			{journal} {Nat. Phys}\ }\textbf {\bibinfo {volume} {1}},\ \bibinfo {pages}
		{111} (\bibinfo {year} {2005})}\BibitemShut {NoStop}%
	\bibitem [{\citenamefont {Schmaljohann}\ \emph {et~al.}(2004)\citenamefont
		{Schmaljohann}, \citenamefont {Erhard}, \citenamefont {Kronj\"ager},
		\citenamefont {Kottke}, \citenamefont {van Staa}, \citenamefont
		{Cacciapuoti}, \citenamefont {Arlt}, \citenamefont {Bongs},\ and\
		\citenamefont {Sengstock}}]{Schmaljohann2004}%
	\BibitemOpen
	\bibfield  {author} {\bibinfo {author} {\bibfnamefont {H.}~\bibnamefont
			{Schmaljohann}}, \bibinfo {author} {\bibfnamefont {M.}~\bibnamefont
			{Erhard}}, \bibinfo {author} {\bibfnamefont {J.}~\bibnamefont {Kronj\"ager}},
		\bibinfo {author} {\bibfnamefont {M.}~\bibnamefont {Kottke}}, \bibinfo
		{author} {\bibfnamefont {S.}~\bibnamefont {van Staa}}, \bibinfo {author}
		{\bibfnamefont {L.}~\bibnamefont {Cacciapuoti}}, \bibinfo {author}
		{\bibfnamefont {J.~J.}\ \bibnamefont {Arlt}}, \bibinfo {author}
		{\bibfnamefont {K.}~\bibnamefont {Bongs}},\ and\ \bibinfo {author}
		{\bibfnamefont {K.}~\bibnamefont {Sengstock}},\ }\bibfield  {title} {\bibinfo
		{title} {{Dynamics of $F=2$ spinor Bose-Einstein condensates}},\ }\href
	{https://doi.org/10.1103/PhysRevLett.92.040402} {\bibfield  {journal}
		{\bibinfo  {journal} {Phys. Rev. Lett.}\ }\textbf {\bibinfo {volume} {92}},\
		\bibinfo {pages} {040402} (\bibinfo {year} {2004})}\BibitemShut {NoStop}%
	\bibitem [{\citenamefont {Al~Khawaja}\ and\ \citenamefont
		{Stoof}(2001)}]{al2001}%
	\BibitemOpen
	\bibfield  {author} {\bibinfo {author} {\bibfnamefont {U.}~\bibnamefont
			{Al~Khawaja}}\ and\ \bibinfo {author} {\bibfnamefont {H.}~\bibnamefont
			{Stoof}},\ }\bibfield  {title} {\bibinfo {title} {{Skyrmions in a
				ferromagnetic Bose-Einstein condensate}},\ }\href
	{https://doi.org/10.1038/35082010} {\bibfield  {journal} {\bibinfo  {journal}
			{Nature (London)}\ }\textbf {\bibinfo {volume} {411}},\ \bibinfo {pages}
		{918} (\bibinfo {year} {2001})}\BibitemShut {NoStop}%
	\bibitem [{\citenamefont {Leslie}\ \emph {et~al.}(2009)\citenamefont {Leslie},
		\citenamefont {Hansen}, \citenamefont {Wright}, \citenamefont {Deutsch},\
		and\ \citenamefont {Bigelow}}]{Leslie2009}%
	\BibitemOpen
	\bibfield  {author} {\bibinfo {author} {\bibfnamefont {L.~S.}\ \bibnamefont
			{Leslie}}, \bibinfo {author} {\bibfnamefont {A.}~\bibnamefont {Hansen}},
		\bibinfo {author} {\bibfnamefont {K.~C.}\ \bibnamefont {Wright}}, \bibinfo
		{author} {\bibfnamefont {B.~M.}\ \bibnamefont {Deutsch}},\ and\ \bibinfo
		{author} {\bibfnamefont {N.~P.}\ \bibnamefont {Bigelow}},\ }\bibfield
	{title} {\bibinfo {title} {{Creation and detection of skyrmions in a
				Bose-Einstein condensate}},\ }\href
	{https://doi.org/10.1103/PhysRevLett.103.250401} {\bibfield  {journal}
		{\bibinfo  {journal} {Phys. Rev. Lett.}\ }\textbf {\bibinfo {volume} {103}},\
		\bibinfo {pages} {250401} (\bibinfo {year} {2009})}\BibitemShut {NoStop}%
	\bibitem [{\citenamefont {Choi}\ \emph {et~al.}(2012)\citenamefont {Choi},
		\citenamefont {Kwon},\ and\ \citenamefont {Shin}}]{Choi2012}%
	\BibitemOpen
	\bibfield  {author} {\bibinfo {author} {\bibfnamefont {J.-Y.}\ \bibnamefont
			{Choi}}, \bibinfo {author} {\bibfnamefont {W.~J.}\ \bibnamefont {Kwon}},\
		and\ \bibinfo {author} {\bibfnamefont {Y.-I.}\ \bibnamefont {Shin}},\
	}\bibfield  {title} {\bibinfo {title} {{Observation of topologically stable
				2D skyrmions in an antiferromagnetic spinor Bose-Einstein condensate}},\
	}\href {https://doi.org/10.1103/PhysRevLett.108.035301} {\bibfield  {journal}
		{\bibinfo  {journal} {Phys. Rev. Lett.}\ }\textbf {\bibinfo {volume} {108}},\
		\bibinfo {pages} {035301} (\bibinfo {year} {2012})}\BibitemShut {NoStop}%
	\bibitem [{\citenamefont {Ray}\ \emph {et~al.}(2014)\citenamefont {Ray},
		\citenamefont {Ruokokoski}, \citenamefont {Kandel}, \citenamefont
		{M{\"o}tt{\"o}nen},\ and\ \citenamefont {Hall}}]{ray2014}%
	\BibitemOpen
	\bibfield  {author} {\bibinfo {author} {\bibfnamefont {M.~W.}\ \bibnamefont
			{Ray}}, \bibinfo {author} {\bibfnamefont {E.}~\bibnamefont {Ruokokoski}},
		\bibinfo {author} {\bibfnamefont {S.}~\bibnamefont {Kandel}}, \bibinfo
		{author} {\bibfnamefont {M.}~\bibnamefont {M{\"o}tt{\"o}nen}},\ and\ \bibinfo
		{author} {\bibfnamefont {D.~S.}\ \bibnamefont {Hall}},\ }\bibfield  {title}
	{\bibinfo {title} {{Observation of dirac monopoles in a synthetic magnetic
				field}},\ }\href {https://doi.org/10.1038/nature12954} {\bibfield  {journal}
		{\bibinfo  {journal} {Nature (London)}\ }\textbf {\bibinfo {volume} {505}},\
		\bibinfo {pages} {657} (\bibinfo {year} {2014})}\BibitemShut {NoStop}%
	\bibitem [{\citenamefont {Seo}\ \emph {et~al.}(2015)\citenamefont {Seo},
		\citenamefont {Kang}, \citenamefont {Kwon},\ and\ \citenamefont
		{Shin}}]{Seo2015}%
	\BibitemOpen
	\bibfield  {author} {\bibinfo {author} {\bibfnamefont {S.~W.}\ \bibnamefont
			{Seo}}, \bibinfo {author} {\bibfnamefont {S.}~\bibnamefont {Kang}}, \bibinfo
		{author} {\bibfnamefont {W.~J.}\ \bibnamefont {Kwon}},\ and\ \bibinfo
		{author} {\bibfnamefont {Y.-I.}\ \bibnamefont {Shin}},\ }\bibfield  {title}
	{\bibinfo {title} {{Half-quantum vortices in an antiferromagnetic spinor
				Bose-Einstein condensate}},\ }\href
	{https://doi.org/10.1103/PhysRevLett.115.015301} {\bibfield  {journal}
		{\bibinfo  {journal} {Phys. Rev. Lett.}\ }\textbf {\bibinfo {volume} {115}},\
		\bibinfo {pages} {015301} (\bibinfo {year} {2015})}\BibitemShut {NoStop}%
	\bibitem [{\citenamefont {Hall}\ \emph {et~al.}(2016)\citenamefont {Hall},
		\citenamefont {Ray}, \citenamefont {Tiurev}, \citenamefont {Ruokokoski},
		\citenamefont {Gheorghe},\ and\ \citenamefont
		{M{\"o}tt{\"o}nen}}]{hall2016tying}%
	\BibitemOpen
	\bibfield  {author} {\bibinfo {author} {\bibfnamefont {D.~S.}\ \bibnamefont
			{Hall}}, \bibinfo {author} {\bibfnamefont {M.~W.}\ \bibnamefont {Ray}},
		\bibinfo {author} {\bibfnamefont {K.}~\bibnamefont {Tiurev}}, \bibinfo
		{author} {\bibfnamefont {E.}~\bibnamefont {Ruokokoski}}, \bibinfo {author}
		{\bibfnamefont {A.~H.}\ \bibnamefont {Gheorghe}},\ and\ \bibinfo {author}
		{\bibfnamefont {M.}~\bibnamefont {M{\"o}tt{\"o}nen}},\ }\bibfield  {title}
	{\bibinfo {title} {{Tying quantum knots}},\ }\href
	{https://doi.org/10.1038/nphys3624} {\bibfield  {journal} {\bibinfo
			{journal} {Nat. Phys.}\ }\textbf {\bibinfo {volume} {12}},\ \bibinfo {pages}
		{478} (\bibinfo {year} {2016})}\BibitemShut {NoStop}%
	\bibitem [{\citenamefont {Sadler}\ \emph {et~al.}(2006)\citenamefont {Sadler},
		\citenamefont {Higbie}, \citenamefont {Leslie}, \citenamefont
		{Vengalattore},\ and\ \citenamefont {Stamper-Kurn}}]{sadler2006}%
	\BibitemOpen
	\bibfield  {author} {\bibinfo {author} {\bibfnamefont {L.~E.}\ \bibnamefont
			{Sadler}}, \bibinfo {author} {\bibfnamefont {J.~M.}\ \bibnamefont {Higbie}},
		\bibinfo {author} {\bibfnamefont {S.~R.}\ \bibnamefont {Leslie}}, \bibinfo
		{author} {\bibfnamefont {M.}~\bibnamefont {Vengalattore}},\ and\ \bibinfo
		{author} {\bibfnamefont {D.~M.}\ \bibnamefont {Stamper-Kurn}},\ }\bibfield
	{title} {\bibinfo {title} {{Spontaneous symmetry breaking in a quenched
				ferromagnetic spinor Bose-Einstein condensate}},\ }\href
	{https://doi.org/10.1038/nature05094} {\bibfield  {journal} {\bibinfo
			{journal} {Nature (London)}\ }\textbf {\bibinfo {volume} {443}},\ \bibinfo
		{pages} {312} (\bibinfo {year} {2006})}\BibitemShut {NoStop}%
	\bibitem [{\citenamefont {Eto}\ \emph {et~al.}(2014)\citenamefont {Eto},
		\citenamefont {Saito},\ and\ \citenamefont {Hirano}}]{Eto2014}%
	\BibitemOpen
	\bibfield  {author} {\bibinfo {author} {\bibfnamefont {Y.}~\bibnamefont
			{Eto}}, \bibinfo {author} {\bibfnamefont {H.}~\bibnamefont {Saito}},\ and\
		\bibinfo {author} {\bibfnamefont {T.}~\bibnamefont {Hirano}},\ }\bibfield
	{title} {\bibinfo {title} {{Observation of dipole-induced spin texture in an
				$^{87}\mathrm{Rb}$ Bose-Einstein condensate}},\ }\href
	{https://doi.org/10.1103/PhysRevLett.112.185301} {\bibfield  {journal}
		{\bibinfo  {journal} {Phys. Rev. Lett.}\ }\textbf {\bibinfo {volume} {112}},\
		\bibinfo {pages} {185301} (\bibinfo {year} {2014})}\BibitemShut {NoStop}%
	\bibitem [{\citenamefont {Ciobanu}\ \emph {et~al.}(2000)\citenamefont
		{Ciobanu}, \citenamefont {Yip},\ and\ \citenamefont {Ho}}]{Ciobanu2000}%
	\BibitemOpen
	\bibfield  {author} {\bibinfo {author} {\bibfnamefont {C.~V.}\ \bibnamefont
			{Ciobanu}}, \bibinfo {author} {\bibfnamefont {S.-K.}\ \bibnamefont {Yip}},\
		and\ \bibinfo {author} {\bibfnamefont {T.-L.}\ \bibnamefont {Ho}},\
	}\bibfield  {title} {\bibinfo {title} {{Phase diagrams of $F=2$ spinor
				Bose-Einstein condensates}},\ }\href
	{https://doi.org/10.1103/PhysRevA.61.033607} {\bibfield  {journal} {\bibinfo
			{journal} {Phys. Rev. A}\ }\textbf {\bibinfo {volume} {61}},\ \bibinfo
		{pages} {033607} (\bibinfo {year} {2000})}\BibitemShut {NoStop}%
	\bibitem [{\citenamefont {Koashi}\ and\ \citenamefont
		{Ueda}(2000)}]{Koashi2000}%
	\BibitemOpen
	\bibfield  {author} {\bibinfo {author} {\bibfnamefont {M.}~\bibnamefont
			{Koashi}}\ and\ \bibinfo {author} {\bibfnamefont {M.}~\bibnamefont {Ueda}},\
	}\bibfield  {title} {\bibinfo {title} {{Exact eigenstates and magnetic
				response of spin-1 and spin-2 Bose-Einstein condensates}},\ }\href
	{https://doi.org/10.1103/PhysRevLett.84.1066} {\bibfield  {journal} {\bibinfo
			{journal} {Phys. Rev. Lett.}\ }\textbf {\bibinfo {volume} {84}},\ \bibinfo
		{pages} {1066} (\bibinfo {year} {2000})}\BibitemShut {NoStop}%
	\bibitem [{\citenamefont {Ueda}\ and\ \citenamefont {Koashi}(2002)}]{Ueda2002}%
	\BibitemOpen
	\bibfield  {author} {\bibinfo {author} {\bibfnamefont {M.}~\bibnamefont
			{Ueda}}\ and\ \bibinfo {author} {\bibfnamefont {M.}~\bibnamefont {Koashi}},\
	}\bibfield  {title} {\bibinfo {title} {{Theory of spin-2 Bose-Einstein
				condensates: Spin correlations, magnetic response, and excitation spectra}},\
	}\href {https://doi.org/10.1103/PhysRevA.65.063602} {\bibfield  {journal}
		{\bibinfo  {journal} {Phys. Rev. A}\ }\textbf {\bibinfo {volume} {65}},\
		\bibinfo {pages} {063602} (\bibinfo {year} {2002})}\BibitemShut {NoStop}%
	\bibitem [{\citenamefont {G\"orlitz}\ \emph {et~al.}(2003)\citenamefont
		{G\"orlitz}, \citenamefont {Gustavson}, \citenamefont {Leanhardt},
		\citenamefont {L\"ow}, \citenamefont {Chikkatur}, \citenamefont {Gupta},
		\citenamefont {Inouye}, \citenamefont {Pritchard},\ and\ \citenamefont
		{Ketterle}}]{gorlitz2003}%
	\BibitemOpen
	\bibfield  {author} {\bibinfo {author} {\bibfnamefont {A.}~\bibnamefont
			{G\"orlitz}}, \bibinfo {author} {\bibfnamefont {T.~L.}\ \bibnamefont
			{Gustavson}}, \bibinfo {author} {\bibfnamefont {A.~E.}\ \bibnamefont
			{Leanhardt}}, \bibinfo {author} {\bibfnamefont {R.}~\bibnamefont {L\"ow}},
		\bibinfo {author} {\bibfnamefont {A.~P.}\ \bibnamefont {Chikkatur}}, \bibinfo
		{author} {\bibfnamefont {S.}~\bibnamefont {Gupta}}, \bibinfo {author}
		{\bibfnamefont {S.}~\bibnamefont {Inouye}}, \bibinfo {author} {\bibfnamefont
			{D.~E.}\ \bibnamefont {Pritchard}},\ and\ \bibinfo {author} {\bibfnamefont
			{W.}~\bibnamefont {Ketterle}},\ }\bibfield  {title} {\bibinfo {title}
		{{Sodium Bose-Einstein condensates in the $F=2$ state in a large-volume
				optical trap}},\ }\href {https://doi.org/10.1103/PhysRevLett.90.090401}
	{\bibfield  {journal} {\bibinfo  {journal} {Phys. Rev. Lett.}\ }\textbf
		{\bibinfo {volume} {90}},\ \bibinfo {pages} {090401} (\bibinfo {year}
		{2003})}\BibitemShut {NoStop}%
	\bibitem [{\citenamefont {Kuwamoto}\ \emph {et~al.}(2004)\citenamefont
		{Kuwamoto}, \citenamefont {Araki}, \citenamefont {Eno},\ and\ \citenamefont
		{Hirano}}]{Kuwamoto2004}%
	\BibitemOpen
	\bibfield  {author} {\bibinfo {author} {\bibfnamefont {T.}~\bibnamefont
			{Kuwamoto}}, \bibinfo {author} {\bibfnamefont {K.}~\bibnamefont {Araki}},
		\bibinfo {author} {\bibfnamefont {T.}~\bibnamefont {Eno}},\ and\ \bibinfo
		{author} {\bibfnamefont {T.}~\bibnamefont {Hirano}},\ }\bibfield  {title}
	{\bibinfo {title} {{Magnetic field dependence of the dynamics of
				$^{87}\mathrm{Rb}$ spin-2 Bose-Einstein condensates}},\ }\href
	{https://doi.org/10.1103/PhysRevA.69.063604} {\bibfield  {journal} {\bibinfo
			{journal} {Phys. Rev. A}\ }\textbf {\bibinfo {volume} {69}},\ \bibinfo
		{pages} {063604} (\bibinfo {year} {2004})}\BibitemShut {NoStop}%
	\bibitem [{\citenamefont {Barnett}\ \emph {et~al.}(2006)\citenamefont
		{Barnett}, \citenamefont {Turner},\ and\ \citenamefont
		{Demler}}]{Barnett2006}%
	\BibitemOpen
	\bibfield  {author} {\bibinfo {author} {\bibfnamefont {R.}~\bibnamefont
			{Barnett}}, \bibinfo {author} {\bibfnamefont {A.}~\bibnamefont {Turner}},\
		and\ \bibinfo {author} {\bibfnamefont {E.}~\bibnamefont {Demler}},\
	}\bibfield  {title} {\bibinfo {title} {{Classifying novel phases of spinor
				atoms}},\ }\href {https://doi.org/10.1103/PhysRevLett.97.180412} {\bibfield
		{journal} {\bibinfo  {journal} {Phys. Rev. Lett.}\ }\textbf {\bibinfo
			{volume} {97}},\ \bibinfo {pages} {180412} (\bibinfo {year}
		{2006})}\BibitemShut {NoStop}%
	\bibitem [{\citenamefont {Diener}\ and\ \citenamefont {Ho}(2006)}]{ho2006}%
	\BibitemOpen
	\bibfield  {author} {\bibinfo {author} {\bibfnamefont {R.~B.}\ \bibnamefont
			{Diener}}\ and\ \bibinfo {author} {\bibfnamefont {T.-L.}\ \bibnamefont
			{Ho}},\ }\bibfield  {title} {\bibinfo {title} {{$^{52}\mathrm{Cr}$ Spinor
				condensate: A biaxial or uniaxial spin nematic}},\ }\href
	{https://doi.org/10.1103/PhysRevLett.96.190405} {\bibfield  {journal}
		{\bibinfo  {journal} {Phys. Rev. Lett.}\ }\textbf {\bibinfo {volume} {96}},\
		\bibinfo {pages} {190405} (\bibinfo {year} {2006})}\BibitemShut {NoStop}%
	\bibitem [{\citenamefont {Santos}\ and\ \citenamefont
		{Pfau}(2006)}]{santos2006}%
	\BibitemOpen
	\bibfield  {author} {\bibinfo {author} {\bibfnamefont {L.}~\bibnamefont
			{Santos}}\ and\ \bibinfo {author} {\bibfnamefont {T.}~\bibnamefont {Pfau}},\
	}\bibfield  {title} {\bibinfo {title} {{Spin-3 chromium Bose-Einstein
				condensates}},\ }\href {https://doi.org/10.1103/PhysRevLett.96.190404}
	{\bibfield  {journal} {\bibinfo  {journal} {Phys. Rev. Lett.}\ }\textbf
		{\bibinfo {volume} {96}},\ \bibinfo {pages} {190404} (\bibinfo {year}
		{2006})}\BibitemShut {NoStop}%
	\bibitem [{\citenamefont {M\"akel\"a}\ and\ \citenamefont
		{Suominen}(2007)}]{makela2007}%
	\BibitemOpen
	\bibfield  {author} {\bibinfo {author} {\bibfnamefont {H.}~\bibnamefont
			{M\"akel\"a}}\ and\ \bibinfo {author} {\bibfnamefont {K.-A.}\ \bibnamefont
			{Suominen}},\ }\bibfield  {title} {\bibinfo {title} {{Ground states of spin-3
				Bose-Einstein condensates for conserved magnetization}},\ }\href
	{https://doi.org/10.1103/PhysRevA.75.033610} {\bibfield  {journal} {\bibinfo
			{journal} {Phys. Rev. A}\ }\textbf {\bibinfo {volume} {75}},\ \bibinfo
		{pages} {033610} (\bibinfo {year} {2007})}\BibitemShut {NoStop}%
	\bibitem [{\citenamefont {Luo}\ \emph {et~al.}(2007)\citenamefont {Luo},
		\citenamefont {Li},\ and\ \citenamefont {Bao}}]{luo2007bose}%
	\BibitemOpen
	\bibfield  {author} {\bibinfo {author} {\bibfnamefont {M.}~\bibnamefont
			{Luo}}, \bibinfo {author} {\bibfnamefont {Z.}~\bibnamefont {Li}},\ and\
		\bibinfo {author} {\bibfnamefont {C.}~\bibnamefont {Bao}},\ }\bibfield
	{title} {\bibinfo {title} {{Bose-Einstein condensate of a mixture of two
				species of spin-1 atoms}},\ }\href
	{https://doi.org/10.1103/PhysRevA.75.043609} {\bibfield  {journal} {\bibinfo
			{journal} {Phys. Rev. A}\ }\textbf {\bibinfo {volume} {75}},\ \bibinfo
		{pages} {043609} (\bibinfo {year} {2007})}\BibitemShut {NoStop}%
	\bibitem [{\citenamefont {Xu}\ \emph {et~al.}(2009)\citenamefont {Xu},
		\citenamefont {Zhang},\ and\ \citenamefont {You}}]{xu2009binary}%
	\BibitemOpen
	\bibfield  {author} {\bibinfo {author} {\bibfnamefont {Z.~F.}\ \bibnamefont
			{Xu}}, \bibinfo {author} {\bibfnamefont {Y.}~\bibnamefont {Zhang}},\ and\
		\bibinfo {author} {\bibfnamefont {L.}~\bibnamefont {You}},\ }\bibfield
	{title} {\bibinfo {title} {{Binary mixture of spinor atomic Bose-Einstein
				condensates}},\ }\href {https://doi.org/10.1103/PhysRevA.79.023613}
	{\bibfield  {journal} {\bibinfo  {journal} {Phys. Rev. A}\ }\textbf {\bibinfo
			{volume} {79}},\ \bibinfo {pages} {023613} (\bibinfo {year}
		{2009})}\BibitemShut {NoStop}%
	\bibitem [{\citenamefont {Shi}(2010)}]{shi2010ground}%
	\BibitemOpen
	\bibfield  {author} {\bibinfo {author} {\bibfnamefont {Y.}~\bibnamefont
			{Shi}},\ }\bibfield  {title} {\bibinfo {title} {{Ground states of a mixture
				of two species of spinor Bose gases with interspecies spin exchange}},\
	}\href {https://doi.org/10.1103/PhysRevA.82.023603} {\bibfield  {journal}
		{\bibinfo  {journal} {Phys. Rev. A}\ }\textbf {\bibinfo {volume} {82}},\
		\bibinfo {pages} {023603} (\bibinfo {year} {2010})}\BibitemShut {NoStop}%
	\bibitem [{\citenamefont {Xu}\ \emph {et~al.}(2010{\natexlab{a}})\citenamefont
		{Xu}, \citenamefont {Zhang}, \citenamefont {Zhang},\ and\ \citenamefont
		{You}}]{xu2010quantum}%
	\BibitemOpen
	\bibfield  {author} {\bibinfo {author} {\bibfnamefont {Z.~F.}\ \bibnamefont
			{Xu}}, \bibinfo {author} {\bibfnamefont {J.}~\bibnamefont {Zhang}}, \bibinfo
		{author} {\bibfnamefont {Y.}~\bibnamefont {Zhang}},\ and\ \bibinfo {author}
		{\bibfnamefont {L.}~\bibnamefont {You}},\ }\bibfield  {title} {\bibinfo
		{title} {{Quantum states of a binary mixture of spinor Bose-Einstein
				condensates}},\ }\href {https://doi.org/10.1103/PhysRevA.81.033603}
	{\bibfield  {journal} {\bibinfo  {journal} {Phys. Rev. A}\ }\textbf {\bibinfo
			{volume} {81}},\ \bibinfo {pages} {033603} (\bibinfo {year}
		{2010}{\natexlab{a}})}\BibitemShut {NoStop}%
	\bibitem [{\citenamefont {Xu}\ \emph {et~al.}(2010{\natexlab{b}})\citenamefont
		{Xu}, \citenamefont {Mei}, \citenamefont {L\"u},\ and\ \citenamefont
		{You}}]{xu2010spontaneously}%
	\BibitemOpen
	\bibfield  {author} {\bibinfo {author} {\bibfnamefont {Z.~F.}\ \bibnamefont
			{Xu}}, \bibinfo {author} {\bibfnamefont {J.~W.}\ \bibnamefont {Mei}},
		\bibinfo {author} {\bibfnamefont {R.}~\bibnamefont {L\"u}},\ and\ \bibinfo
		{author} {\bibfnamefont {L.}~\bibnamefont {You}},\ }\bibfield  {title}
	{\bibinfo {title} {{Spontaneously axisymmetry-breaking phase in a binary
				mixture of spinor Bose-Einstein condensates}},\ }\href
	{https://doi.org/10.1103/PhysRevA.82.053626} {\bibfield  {journal} {\bibinfo
			{journal} {Phys. Rev. A}\ }\textbf {\bibinfo {volume} {82}},\ \bibinfo
		{pages} {053626} (\bibinfo {year} {2010}{\natexlab{b}})}\BibitemShut
	{NoStop}%
	\bibitem [{\citenamefont {Zhang}\ \emph {et~al.}(2010)\citenamefont {Zhang},
		\citenamefont {Xu}, \citenamefont {You},\ and\ \citenamefont
		{Zhang}}]{zhang2010atomic}%
	\BibitemOpen
	\bibfield  {author} {\bibinfo {author} {\bibfnamefont {J.}~\bibnamefont
			{Zhang}}, \bibinfo {author} {\bibfnamefont {Z.~F.}\ \bibnamefont {Xu}},
		\bibinfo {author} {\bibfnamefont {L.}~\bibnamefont {You}},\ and\ \bibinfo
		{author} {\bibfnamefont {Y.}~\bibnamefont {Zhang}},\ }\bibfield  {title}
	{\bibinfo {title} {{Atomic-number fluctuations in a mixture of
				condensates}},\ }\href {https://doi.org/10.1103/PhysRevA.82.013625}
	{\bibfield  {journal} {\bibinfo  {journal} {Phys. Rev. A}\ }\textbf {\bibinfo
			{volume} {82}},\ \bibinfo {pages} {013625} (\bibinfo {year}
		{2010})}\BibitemShut {NoStop}%
	\bibitem [{\citenamefont {Shi}\ and\ \citenamefont {Ge}(2011)}]{shi2011three}%
	\BibitemOpen
	\bibfield  {author} {\bibinfo {author} {\bibfnamefont {Y.}~\bibnamefont
			{Shi}}\ and\ \bibinfo {author} {\bibfnamefont {L.}~\bibnamefont {Ge}},\
	}\bibfield  {title} {\bibinfo {title} {{Three-dimensional quantum phase
				diagram of the exact ground states of a mixture of two species of spin-$1$
				Bose gases with interspecies spin exchange}},\ }\href
	{https://doi.org/10.1103/PhysRevA.83.013616} {\bibfield  {journal} {\bibinfo
			{journal} {Phys. Rev. A}\ }\textbf {\bibinfo {volume} {83}},\ \bibinfo
		{pages} {013616} (\bibinfo {year} {2011})}\BibitemShut {NoStop}%
	\bibitem [{\citenamefont {Zhang}\ \emph {et~al.}(2011)\citenamefont {Zhang},
		\citenamefont {Li},\ and\ \citenamefont {Zhang}}]{zhang2011interspecies}%
	\BibitemOpen
	\bibfield  {author} {\bibinfo {author} {\bibfnamefont {J.}~\bibnamefont
			{Zhang}}, \bibinfo {author} {\bibfnamefont {T.}~\bibnamefont {Li}},\ and\
		\bibinfo {author} {\bibfnamefont {Y.}~\bibnamefont {Zhang}},\ }\bibfield
	{title} {\bibinfo {title} {{Interspecies singlet pairing in a mixture of two
				spin-1 Bose condensates}},\ }\href
	{https://doi.org/10.1103/PhysRevA.83.023614} {\bibfield  {journal} {\bibinfo
			{journal} {Phys. Rev. A}\ }\textbf {\bibinfo {volume} {83}},\ \bibinfo
		{pages} {023614} (\bibinfo {year} {2011})}\BibitemShut {NoStop}%
	\bibitem [{\citenamefont {Xu}\ \emph {et~al.}(2011)\citenamefont {Xu},
		\citenamefont {L\"u},\ and\ \citenamefont {You}}]{xu2011quantum}%
	\BibitemOpen
	\bibfield  {author} {\bibinfo {author} {\bibfnamefont {Z.~F.}\ \bibnamefont
			{Xu}}, \bibinfo {author} {\bibfnamefont {R.}~\bibnamefont {L\"u}},\ and\
		\bibinfo {author} {\bibfnamefont {L.}~\bibnamefont {You}},\ }\bibfield
	{title} {\bibinfo {title} {{Quantum entangled ground states of two spinor
				Bose-Einstein condensates}},\ }\href
	{https://doi.org/10.1103/PhysRevA.84.063634} {\bibfield  {journal} {\bibinfo
			{journal} {Phys. Rev. A}\ }\textbf {\bibinfo {volume} {84}},\ \bibinfo
		{pages} {063634} (\bibinfo {year} {2011})}\BibitemShut {NoStop}%
	\bibitem [{\citenamefont {Xu}\ \emph {et~al.}(2012)\citenamefont {Xu},
		\citenamefont {Wang},\ and\ \citenamefont {You}}]{xu2012quantum}%
	\BibitemOpen
	\bibfield  {author} {\bibinfo {author} {\bibfnamefont {Z.~F.}\ \bibnamefont
			{Xu}}, \bibinfo {author} {\bibfnamefont {D.~J.}\ \bibnamefont {Wang}},\ and\
		\bibinfo {author} {\bibfnamefont {L.}~\bibnamefont {You}},\ }\bibfield
	{title} {\bibinfo {title} {{Quantum spin mixing in a binary mixture of spin-1
				atomic condensates}},\ }\href {https://doi.org/10.1103/PhysRevA.86.013632}
	{\bibfield  {journal} {\bibinfo  {journal} {Phys. Rev. A}\ }\textbf {\bibinfo
			{volume} {86}},\ \bibinfo {pages} {013632} (\bibinfo {year}
		{2012})}\BibitemShut {NoStop}%
	\bibitem [{\citenamefont {Zhang}\ \emph {et~al.}(2015)\citenamefont {Zhang},
		\citenamefont {Hou}, \citenamefont {Chen},\ and\ \citenamefont
		{Zhang}}]{zhang2015fragmentation}%
	\BibitemOpen
	\bibfield  {author} {\bibinfo {author} {\bibfnamefont {J.}~\bibnamefont
			{Zhang}}, \bibinfo {author} {\bibfnamefont {X.}~\bibnamefont {Hou}}, \bibinfo
		{author} {\bibfnamefont {B.}~\bibnamefont {Chen}},\ and\ \bibinfo {author}
		{\bibfnamefont {Y.}~\bibnamefont {Zhang}},\ }\bibfield  {title} {\bibinfo
		{title} {{Fragmentation of a spin-1 mixture in a magnetic field}},\ }\href
	{https://doi.org/10.1103/PhysRevA.91.013628} {\bibfield  {journal} {\bibinfo
			{journal} {Phys. Rev. A}\ }\textbf {\bibinfo {volume} {91}},\ \bibinfo
		{pages} {013628} (\bibinfo {year} {2015})}\BibitemShut {NoStop}%
	\bibitem [{\citenamefont {Chen}\ \emph {et~al.}(2018)\citenamefont {Chen},
		\citenamefont {Xu},\ and\ \citenamefont {You}}]{chen2018resonant}%
	\BibitemOpen
	\bibfield  {author} {\bibinfo {author} {\bibfnamefont {J.-J.}\ \bibnamefont
			{Chen}}, \bibinfo {author} {\bibfnamefont {Z.-F.}\ \bibnamefont {Xu}},\ and\
		\bibinfo {author} {\bibfnamefont {L.}~\bibnamefont {You}},\ }\bibfield
	{title} {\bibinfo {title} {{Resonant spin exchange between heteronuclear
				atoms assisted by periodic driving}},\ }\href
	{https://doi.org/10.1103/PhysRevA.98.023601} {\bibfield  {journal} {\bibinfo
			{journal} {Phys. Rev. A}\ }\textbf {\bibinfo {volume} {98}},\ \bibinfo
		{pages} {023601} (\bibinfo {year} {2018})}\BibitemShut {NoStop}%
	\bibitem [{\citenamefont {He}\ \emph {et~al.}(2019{\natexlab{a}})\citenamefont
		{He}, \citenamefont {Liu},\ and\ \citenamefont {Bao}}]{he2019}%
	\BibitemOpen
	\bibfield  {author} {\bibinfo {author} {\bibfnamefont {Y.~Z.}\ \bibnamefont
			{He}}, \bibinfo {author} {\bibfnamefont {Y.~M.}\ \bibnamefont {Liu}},\ and\
		\bibinfo {author} {\bibfnamefont {C.~G.}\ \bibnamefont {Bao}},\ }\bibfield
	{title} {\bibinfo {title} {{Variation of the spin textures of 2-species
				spin-1 condensates studied beyond the single spatial mode approximation and
				the experimental identification of these textures}},\ }\href
	{https://doi.org/10.1088/1402-4896/ab15fd} {\bibfield  {journal} {\bibinfo
			{journal} {Phys. Scr.}\ }\textbf {\bibinfo {volume} {94}},\ \bibinfo {pages}
		{115403} (\bibinfo {year} {2019}{\natexlab{a}})}\BibitemShut {NoStop}%
	\bibitem [{\citenamefont {He}\ \emph {et~al.}(2019{\natexlab{b}})\citenamefont
		{He}, \citenamefont {Liu},\ and\ \citenamefont {Bao}}]{he2019_2}%
	\BibitemOpen
	\bibfield  {author} {\bibinfo {author} {\bibfnamefont {Y.~Z.}\ \bibnamefont
			{He}}, \bibinfo {author} {\bibfnamefont {Y.~M.}\ \bibnamefont {Liu}},\ and\
		\bibinfo {author} {\bibfnamefont {C.~G.}\ \bibnamefont {Bao}},\ }\bibfield
	{title} {\bibinfo {title} {{Spin-textures of the condensates with two kinds
				of spin-1 atoms studied beyond the single spatial mode approximation}},\
	}\href {https://doi.org/10.1007/s10909-019-02195-6} {\bibfield  {journal}
		{\bibinfo  {journal} {J. Low Temp. Phys.}\ }\textbf {\bibinfo {volume}
			{196}},\ \bibinfo {pages} {458} (\bibinfo {year}
		{2019}{\natexlab{b}})}\BibitemShut {NoStop}%
	\bibitem [{\citenamefont {Jie}\ \emph {et~al.}(2021)\citenamefont {Jie},
		\citenamefont {Yu}, \citenamefont {Wang},\ and\ \citenamefont
		{Zhang}}]{jie2021laser}%
	\BibitemOpen
	\bibfield  {author} {\bibinfo {author} {\bibfnamefont {J.}~\bibnamefont
			{Jie}}, \bibinfo {author} {\bibfnamefont {Y.}~\bibnamefont {Yu}}, \bibinfo
		{author} {\bibfnamefont {D.}~\bibnamefont {Wang}},\ and\ \bibinfo {author}
		{\bibfnamefont {P.}~\bibnamefont {Zhang}},\ }\bibfield  {title} {\bibinfo
		{title} {{Laser control of the singlet-pairing process in an ultracold spinor
				mixture}},\ }\href {https://doi.org/10.1103/PhysRevA.103.053321} {\bibfield
		{journal} {\bibinfo  {journal} {Phys. Rev. A}\ }\textbf {\bibinfo {volume}
			{103}},\ \bibinfo {pages} {053321} (\bibinfo {year} {2021})}\BibitemShut
	{NoStop}%
	\bibitem [{\citenamefont {Li}\ \emph {et~al.}(2015)\citenamefont {Li},
		\citenamefont {Zhu}, \citenamefont {He}, \citenamefont {Wang}, \citenamefont
		{Guo}, \citenamefont {Xu}, \citenamefont {Zhang},\ and\ \citenamefont
		{Wang}}]{li2015coherent}%
	\BibitemOpen
	\bibfield  {author} {\bibinfo {author} {\bibfnamefont {X.}~\bibnamefont
			{Li}}, \bibinfo {author} {\bibfnamefont {B.}~\bibnamefont {Zhu}}, \bibinfo
		{author} {\bibfnamefont {X.}~\bibnamefont {He}}, \bibinfo {author}
		{\bibfnamefont {F.}~\bibnamefont {Wang}}, \bibinfo {author} {\bibfnamefont
			{M.}~\bibnamefont {Guo}}, \bibinfo {author} {\bibfnamefont {Z.-F.}\
			\bibnamefont {Xu}}, \bibinfo {author} {\bibfnamefont {S.}~\bibnamefont
			{Zhang}},\ and\ \bibinfo {author} {\bibfnamefont {D.}~\bibnamefont {Wang}},\
	}\bibfield  {title} {\bibinfo {title} {{Coherent heteronuclear spin dynamics
				in an ultracold spinor mixture}},\ }\href
	{https://doi.org/10.1103/PhysRevLett.114.255301} {\bibfield  {journal}
		{\bibinfo  {journal} {Phys. Rev. Lett.}\ }\textbf {\bibinfo {volume} {114}},\
		\bibinfo {pages} {255301} (\bibinfo {year} {2015})}\BibitemShut {NoStop}%
	\bibitem [{\citenamefont {Liu}\ \emph {et~al.}(2017)\citenamefont {Liu},
		\citenamefont {He},\ and\ \citenamefont {Bao}}]{liu2017}%
	\BibitemOpen
	\bibfield  {author} {\bibinfo {author} {\bibfnamefont {Y.~M.}\ \bibnamefont
			{Liu}}, \bibinfo {author} {\bibfnamefont {Y.~Z.}\ \bibnamefont {He}},\ and\
		\bibinfo {author} {\bibfnamefont {C.~G.}\ \bibnamefont {Bao}},\ }\bibfield
	{title} {\bibinfo {title} {{Singularity in the matrix of the coupled
				Gross-Pitaevskii equations and the related state-transitions in three-species
				condensates}},\ }\href {https://doi.org/10.1038/s41598-017-06843-3}
	{\bibfield  {journal} {\bibinfo  {journal} {Sci. Rep.}\ }\textbf {\bibinfo
			{volume} {7}},\ \bibinfo {pages} {6585} (\bibinfo {year} {2017})}\BibitemShut
	{NoStop}%
	\bibitem [{\citenamefont {He}\ \emph {et~al.}(2020)\citenamefont {He},
		\citenamefont {Liu},\ and\ \citenamefont {Bao}}]{he2020}%
	\BibitemOpen
	\bibfield  {author} {\bibinfo {author} {\bibfnamefont {Y.~Z.}\ \bibnamefont
			{He}}, \bibinfo {author} {\bibfnamefont {Y.~M.}\ \bibnamefont {Liu}},\ and\
		\bibinfo {author} {\bibfnamefont {C.~G.}\ \bibnamefont {Bao}},\ }\bibfield
	{title} {\bibinfo {title} {{Spin-structures of the Bose-Einstein condensates
				with three kinds of spin-1 atoms}},\ }\href
	{https://doi.org/10.1038/s41598-020-59540-z} {\bibfield  {journal} {\bibinfo
			{journal} {Sci. Rep.}\ }\textbf {\bibinfo {volume} {10}},\ \bibinfo {pages}
		{2727} (\bibinfo {year} {2020})}\BibitemShut {NoStop}%
	\bibitem [{\citenamefont {He}\ \emph {et~al.}(2022)\citenamefont {He},
		\citenamefont {Liu},\ and\ \citenamefont {Bao}}]{he2022}%
	\BibitemOpen
	\bibfield  {author} {\bibinfo {author} {\bibfnamefont {Y.~Z.}\ \bibnamefont
			{He}}, \bibinfo {author} {\bibfnamefont {Y.~M.}\ \bibnamefont {Liu}},\ and\
		\bibinfo {author} {\bibfnamefont {C.~G.}\ \bibnamefont {Bao}},\ }\bibfield
	{title} {\bibinfo {title} {{Effect of the singlet pairing force on the spin
				structures of 3-species Bose-Einstein condensates with spin-1 atoms}},\
	}\href {https://doi.org/10.1007/s10909-021-02647-y} {\bibfield  {journal}
		{\bibinfo  {journal} {J. Low Temp. Phys.}\ }\textbf {\bibinfo {volume}
			{206}},\ \bibinfo {pages} {167} (\bibinfo {year} {2022})}\BibitemShut
	{NoStop}%
	\bibitem [{\citenamefont {Irikura}\ \emph {et~al.}(2018)\citenamefont
		{Irikura}, \citenamefont {Eto}, \citenamefont {Hirano},\ and\ \citenamefont
		{Saito}}]{irikura2018}%
	\BibitemOpen
	\bibfield  {author} {\bibinfo {author} {\bibfnamefont {N.}~\bibnamefont
			{Irikura}}, \bibinfo {author} {\bibfnamefont {Y.}~\bibnamefont {Eto}},
		\bibinfo {author} {\bibfnamefont {T.}~\bibnamefont {Hirano}},\ and\ \bibinfo
		{author} {\bibfnamefont {H.}~\bibnamefont {Saito}},\ }\bibfield  {title}
	{\bibinfo {title} {{Ground-state phases of a mixture of spin-1 and spin-2
				Bose-Einstein condensates}},\ }\href
	{https://doi.org/10.1103/PhysRevA.97.023622} {\bibfield  {journal} {\bibinfo
			{journal} {Phys. Rev. A}\ }\textbf {\bibinfo {volume} {97}},\ \bibinfo
		{pages} {023622} (\bibinfo {year} {2018})}\BibitemShut {NoStop}%
	\bibitem [{\citenamefont {Eto}\ \emph {et~al.}(2018)\citenamefont {Eto},
		\citenamefont {Shibayama}, \citenamefont {Saito},\ and\ \citenamefont
		{Hirano}}]{eto2018}%
	\BibitemOpen
	\bibfield  {author} {\bibinfo {author} {\bibfnamefont {Y.}~\bibnamefont
			{Eto}}, \bibinfo {author} {\bibfnamefont {H.}~\bibnamefont {Shibayama}},
		\bibinfo {author} {\bibfnamefont {H.}~\bibnamefont {Saito}},\ and\ \bibinfo
		{author} {\bibfnamefont {T.}~\bibnamefont {Hirano}},\ }\bibfield  {title}
	{\bibinfo {title} {{Spinor dynamics in a mixture of spin-1 and spin-2
				Bose-Einstein condensates}},\ }\href
	{https://doi.org/10.1103/PhysRevA.97.021602} {\bibfield  {journal} {\bibinfo
			{journal} {Phys. Rev. A}\ }\textbf {\bibinfo {volume} {97}},\ \bibinfo
		{pages} {021602(R)} (\bibinfo {year} {2018})}\BibitemShut {NoStop}%
	\bibitem [{\citenamefont {Klausen}\ \emph {et~al.}(2001)\citenamefont
		{Klausen}, \citenamefont {Bohn},\ and\ \citenamefont {Greene}}]{klausen2001}%
	\BibitemOpen
	\bibfield  {author} {\bibinfo {author} {\bibfnamefont {N.~N.}\ \bibnamefont
			{Klausen}}, \bibinfo {author} {\bibfnamefont {J.~L.}\ \bibnamefont {Bohn}},\
		and\ \bibinfo {author} {\bibfnamefont {C.~H.}\ \bibnamefont {Greene}},\
	}\bibfield  {title} {\bibinfo {title} {{Nature of spinor Bose-Einstein
				condensates in rubidium}},\ }\href
	{https://doi.org/10.1103/PhysRevA.64.053602} {\bibfield  {journal} {\bibinfo
			{journal} {Phys. Rev. A}\ }\textbf {\bibinfo {volume} {64}},\ \bibinfo
		{pages} {053602} (\bibinfo {year} {2001})}\BibitemShut {NoStop}%
	\bibitem [{\citenamefont {Widera}\ \emph {et~al.}(2006)\citenamefont {Widera},
		\citenamefont {Gerbier}, \citenamefont {F{\"o}lling}, \citenamefont
		{Gericke}, \citenamefont {Mandel},\ and\ \citenamefont {Bloch}}]{widera2006}%
	\BibitemOpen
	\bibfield  {author} {\bibinfo {author} {\bibfnamefont {A.}~\bibnamefont
			{Widera}}, \bibinfo {author} {\bibfnamefont {F.}~\bibnamefont {Gerbier}},
		\bibinfo {author} {\bibfnamefont {S.}~\bibnamefont {F{\"o}lling}}, \bibinfo
		{author} {\bibfnamefont {T.}~\bibnamefont {Gericke}}, \bibinfo {author}
		{\bibfnamefont {O.}~\bibnamefont {Mandel}},\ and\ \bibinfo {author}
		{\bibfnamefont {I.}~\bibnamefont {Bloch}},\ }\bibfield  {title} {\bibinfo
		{title} {{Precision measurement of spin-dependent interaction strengths for
				spin-1 and spin-2 $^{87}\mathrm{Rb}$ atoms}},\ }\href
	{https://doi.org/10.1088/1367-2630/8/8/152} {\bibfield  {journal} {\bibinfo
			{journal} {New J. Phys.}\ }\textbf {\bibinfo {volume} {8}},\ \bibinfo {pages}
		{152} (\bibinfo {year} {2006})}\BibitemShut {NoStop}%
	\bibitem [{\citenamefont {Egorov}\ \emph {et~al.}(2013)\citenamefont {Egorov},
		\citenamefont {Opanchuk}, \citenamefont {Drummond}, \citenamefont {Hall},
		\citenamefont {Hannaford},\ and\ \citenamefont {Sidorov}}]{egorov2013}%
	\BibitemOpen
	\bibfield  {author} {\bibinfo {author} {\bibfnamefont {M.}~\bibnamefont
			{Egorov}}, \bibinfo {author} {\bibfnamefont {B.}~\bibnamefont {Opanchuk}},
		\bibinfo {author} {\bibfnamefont {P.}~\bibnamefont {Drummond}}, \bibinfo
		{author} {\bibfnamefont {B.~V.}\ \bibnamefont {Hall}}, \bibinfo {author}
		{\bibfnamefont {P.}~\bibnamefont {Hannaford}},\ and\ \bibinfo {author}
		{\bibfnamefont {A.~I.}\ \bibnamefont {Sidorov}},\ }\bibfield  {title}
	{\bibinfo {title} {{Measurement of $s$-wave scattering lengths in a
				two-component Bose-Einstein condensate}},\ }\href
	{https://doi.org/10.1103/PhysRevA.87.053614} {\bibfield  {journal} {\bibinfo
			{journal} {Phys. Rev. A}\ }\textbf {\bibinfo {volume} {87}},\ \bibinfo
		{pages} {053614} (\bibinfo {year} {2013})}\BibitemShut {NoStop}%
	\bibitem [{\citenamefont {Gomez}\ \emph {et~al.}(2019)\citenamefont {Gomez},
		\citenamefont {Mazzinghi}, \citenamefont {Martin}, \citenamefont {Coop},
		\citenamefont {Palacios},\ and\ \citenamefont {Mitchell}}]{gomez2019}%
	\BibitemOpen
	\bibfield  {author} {\bibinfo {author} {\bibfnamefont {P.}~\bibnamefont
			{Gomez}}, \bibinfo {author} {\bibfnamefont {C.}~\bibnamefont {Mazzinghi}},
		\bibinfo {author} {\bibfnamefont {F.}~\bibnamefont {Martin}}, \bibinfo
		{author} {\bibfnamefont {S.}~\bibnamefont {Coop}}, \bibinfo {author}
		{\bibfnamefont {S.}~\bibnamefont {Palacios}},\ and\ \bibinfo {author}
		{\bibfnamefont {M.~W.}\ \bibnamefont {Mitchell}},\ }\bibfield  {title}
	{\bibinfo {title} {{Interferometric measurement of interhyperfine scattering
				lengths in $^{87}\mathrm{Rb}$}},\ }\href
	{https://doi.org/10.1103/PhysRevA.100.032704} {\bibfield  {journal} {\bibinfo
			{journal} {Phys. Rev. A}\ }\textbf {\bibinfo {volume} {100}},\ \bibinfo
		{pages} {032704} (\bibinfo {year} {2019})}\BibitemShut {NoStop}%
	\bibitem [{\citenamefont {Grimm}\ \emph {et~al.}(2000)\citenamefont {Grimm},
		\citenamefont {Weidemüller},\ and\ \citenamefont {Ovchinnikov}}]{Grimm2000}%
	\BibitemOpen
	\bibfield  {author} {\bibinfo {author} {\bibfnamefont {R.}~\bibnamefont
			{Grimm}}, \bibinfo {author} {\bibfnamefont {M.}~\bibnamefont
			{Weidemüller}},\ and\ \bibinfo {author} {\bibfnamefont {Y.~B.}\ \bibnamefont
			{Ovchinnikov}},\ }\bibfield  {title} {\bibinfo {title} {Optical dipole traps
			for neutral atoms},\ }\href
	{https://doi.org/https://doi.org/10.1016/S1049-250X(08)60186-X} {\bibfield
		{journal} {\bibinfo  {journal} {Adv. At. Mol. Opt. Phys.}\ }\textbf {\bibinfo
			{volume} {42}},\ \bibinfo {pages} {95} (\bibinfo {year} {2000})}\BibitemShut
	{NoStop}%
	\bibitem [{\citenamefont {Chiba}\ and\ \citenamefont
		{Saito}(2008)}]{Chiba2008}%
	\BibitemOpen
	\bibfield  {author} {\bibinfo {author} {\bibfnamefont {H.}~\bibnamefont
			{Chiba}}\ and\ \bibinfo {author} {\bibfnamefont {H.}~\bibnamefont {Saito}},\
	}\bibfield  {title} {\bibinfo {title} {{Spin-vortex nucleation in a
				Bose-Einstein condensate by a spin-dependent rotating trap}},\ }\href
	{https://doi.org/10.1103/PhysRevA.78.043602} {\bibfield  {journal} {\bibinfo
			{journal} {Phys. Rev. A}\ }\textbf {\bibinfo {volume} {78}},\ \bibinfo
		{pages} {043602} (\bibinfo {year} {2008})}\BibitemShut {NoStop}%
	\bibitem [{\citenamefont {Kim}\ \emph {et~al.}(2020)\citenamefont {Kim},
		\citenamefont {Hong},\ and\ \citenamefont {Shin}}]{Kim2020}%
	\BibitemOpen
	\bibfield  {author} {\bibinfo {author} {\bibfnamefont {J.~H.}\ \bibnamefont
			{Kim}}, \bibinfo {author} {\bibfnamefont {D.}~\bibnamefont {Hong}},\ and\
		\bibinfo {author} {\bibfnamefont {Y.}~\bibnamefont {Shin}},\ }\bibfield
	{title} {\bibinfo {title} {Observation of two sound modes in a binary
			superfluid gas},\ }\href {https://doi.org/10.1103/PhysRevA.101.061601}
	{\bibfield  {journal} {\bibinfo  {journal} {Phys. Rev. A}\ }\textbf {\bibinfo
			{volume} {101}},\ \bibinfo {pages} {061601(R)} (\bibinfo {year}
		{2020})}\BibitemShut {NoStop}%
	\bibitem [{SM()}]{SM}%
	\BibitemOpen
	\href@noop {} {}\bibinfo {howpublished} {{See Supplemental Material at
			http://link.aps.org/supplemental/... for movies of the
			dynamics.}}\BibitemShut {Stop}%
	\bibitem [{\citenamefont {Tojo}\ \emph {et~al.}(2009)\citenamefont {Tojo},
		\citenamefont {Hayashi}, \citenamefont {Tanabe}, \citenamefont {Hirano},
		\citenamefont {Kawaguchi}, \citenamefont {Saito},\ and\ \citenamefont
		{Ueda}}]{tojo2009spin}%
	\BibitemOpen
	\bibfield  {author} {\bibinfo {author} {\bibfnamefont {S.}~\bibnamefont
			{Tojo}}, \bibinfo {author} {\bibfnamefont {T.}~\bibnamefont {Hayashi}},
		\bibinfo {author} {\bibfnamefont {T.}~\bibnamefont {Tanabe}}, \bibinfo
		{author} {\bibfnamefont {T.}~\bibnamefont {Hirano}}, \bibinfo {author}
		{\bibfnamefont {Y.}~\bibnamefont {Kawaguchi}}, \bibinfo {author}
		{\bibfnamefont {H.}~\bibnamefont {Saito}},\ and\ \bibinfo {author}
		{\bibfnamefont {M.}~\bibnamefont {Ueda}},\ }\bibfield  {title} {\bibinfo
		{title} {{Spin-dependent inelastic collisions in spin-2 Bose-Einstein
				condensates}},\ }\href {https://doi.org/10.1103/PhysRevA.80.042704}
	{\bibfield  {journal} {\bibinfo  {journal} {Phys. Rev. A}\ }\textbf {\bibinfo
			{volume} {80}},\ \bibinfo {pages} {042704} (\bibinfo {year}
		{2009})}\BibitemShut {NoStop}%
	\bibitem [{\citenamefont {Eto}\ \emph {et~al.}(2016)\citenamefont {Eto},
		\citenamefont {Takahashi}, \citenamefont {Kunimi}, \citenamefont {Saito},\
		and\ \citenamefont {Hirano}}]{Eto2016}%
	\BibitemOpen
	\bibfield  {author} {\bibinfo {author} {\bibfnamefont {Y.}~\bibnamefont
			{Eto}}, \bibinfo {author} {\bibfnamefont {M.}~\bibnamefont {Takahashi}},
		\bibinfo {author} {\bibfnamefont {M.}~\bibnamefont {Kunimi}}, \bibinfo
		{author} {\bibfnamefont {H.}~\bibnamefont {Saito}},\ and\ \bibinfo {author}
		{\bibfnamefont {T.}~\bibnamefont {Hirano}},\ }\bibfield  {title} {\bibinfo
		{title} {{Nonequilibrium dynamics induced by miscible–immiscible transition
				in binary Bose–Einstein condensates}},\ }\href
	{https://doi.org/10.1088/1367-2630/18/7/073029} {\bibfield  {journal}
		{\bibinfo  {journal} {New J. Phys.}\ }\textbf {\bibinfo {volume} {18}},\
		\bibinfo {pages} {073029} (\bibinfo {year} {2016})}\BibitemShut {NoStop}%
	\bibitem [{ref()}]{ref1}%
	\BibitemOpen
	\href@noop {} {\bibinfo {title} {{In Fig.~3(e) of Ref.~\cite{Eto2016}, the
				dynamics of a $|1, 0\rangle$ and $|2, -2\rangle$ mixture is shown, which
				should have the same loss rate as the PB state regarding inelastic collisions
				between spin-1 and spin-2. Since no significant atomic loss is observed in
				this dynamics for 600 ms, we can conclude that the relevant loss rate is much
				smaller than that for inelastic collisions between spin-2 atoms, which
				restricts the lifetime of the spin-2 BEC up to $\sim 100$
				ms~\cite{tojo2009spin}}}}\BibitemShut {NoStop}%
\end{thebibliography}
%

\end{document}